\documentclass{emulateapj}

\newcommand{\cena}{NGC~5128}
\newcommand{\etal}{et~al.\ }

\newcommand{\Hb}{H$\beta$}       
\newcommand{\Hga}{H$\gamma_A$}       
\newcommand{\kms}{km~s$^{-1}$}

\newcommand{\uv}{$U$--$V$}
\newcommand{\vi}{$V$--$I$}

\newcommand{\mgfe}{$[MgFe]^{\prime}$}

\submitted{Submitted 2003 August 14; accepted 2003 November 7}
\journalinfo{Astrophysical Journal, v602, 2004 February 20}

\shorttitle{Globular Cluster System of NGC 5128: Ages, Metallicities,
Kinematics, Formation}
\shortauthors{Peng, Ford, \& Freeman}

\begin{document}


\title{The Globular Cluster System of NGC 5128 II. Ages, Metallicities,
Kinematics, and Formation} 


\author{Eric W.\ Peng\altaffilmark{1,2}, Holland C.\ Ford\altaffilmark{1,3}}
\affil{Department of Physics and Astronomy, Johns Hopkins
        University, Baltimore, MD, 21218, USA}
\email{ericpeng@pha.jhu.edu, ford@pha.jhu.edu}

\and

\author{Kenneth C.\ Freeman}
\affil{RSAA, Australian National University, Canberra, ACT, Australia}
\email{kcf@mso.anu.edu.au}


\altaffiltext{1}{Visiting Astronomer, Cerro Tololo Inter-American Observatory,
which is operated by the Association of Universities for Research in
Astronomy, Inc.\ (AURA) under cooperative agreement with the National Science
Foundation.}
\altaffiltext{2}{Current address: 136 Frelinghuysen Road, Physics and
Astronomy, Rutgers University, Piscataway, NJ 08854, USA;
ericpeng@physics.rutgers.edu}
\altaffiltext{3}{Space Telescope Science Institute, 3700 San Martin Drive,
        Baltimore, MD 21218, USA}


\begin{abstract}
We present a study of the nearby post-merger giant elliptical
galaxy, \cena\ (Centaurus A), in which we use the properties of its
globular cluster (GC) and planetary nebula (PN) systems to constrain its
evolution.  Using photometric and spectroscopic data for 215 GCs
presented in Paper I, we study trends in age, metallicity, and
kinematics for the GC system.  We confirm that the GC metallicity
distribution is bimodal, and show that these two sub-populations have
different properties.  Using spectral line index measurements of the
brightest clusters, the metal-poor GCs have old ages like
the Milky Way globular clusters, while the metal-rich GCs have \Hb\
line-strengths that could be interpreted as a mean
age of $\sim5^{+3}_{-2}$~Gyr.  Both populations appear to 
have [Mg/Fe] ratios consistent with that of the 
Galactic GC system, although this quantity is not very well-constrained.  The
kinematics of the metal-rich GCs are similar to those of the planetary
nebulae, exhibiting significant rotation about a misaligned axis, while the
metal-poor GCs have a higher velocity dispersion and show a weaker
kinematic correlation with the field stars.  The total gravitating mass
of \cena\ derived from the GCs is in excellent agreement with the value
derived from stellar (PN) kinematics.  We suggest that these and other data
support a picture in which the main body of \cena\ was formed
3--8~Gyr ago by the dissipational merger of two unequal-mass
disk galaxies supplemented by the continual accretion of both gas-rich
and gas-poor satellites.
\end{abstract}




\section{Introduction}

The fossil record of old star clusters that is present in nearly every galaxy
provides a window onto the vigorous epochs of star formation that mark
its evolutionary history.  In nearby galaxies ($D\lesssim20$~Mpc),
where detailed investigations are possible, globular cluster systems can
provide important leverage on the chemical and dynamical history of their
hosts and their environment.  In spheroidal stellar populations such as
elliptical galaxies and bulges, the observed color
distributions of globular clusters (GCs) alone require either a
significant amount of merging, multiple epochs of star formation (Forbes,
Brodie, \& Grillmair 1997), or both (Ashman \& Zepf 1992).  The hierarchical
merging that has been invoked to explain these systems can be
purely dissipationless (C{\^o}t{\'e}, Marzke, \& West 1998), or also
involve gaseous bursts of star formation (Beasley \etal 2002)  

While much has been learned from the metallicity distributions of these
GC systems, relatively little work has been done on their kinematics
and their ages.  This is mainly because such studies are difficult
for a large fraction of nearby galaxies, 
even with the new generation of 6--10 meter telescopes.  
These properties, however, offer to extend the power of GC
systems as galaxy diagnostics by probing the dynamical
states of galactic sub-components, and the time scales on which they
formed.  Recently, there is evidence that while metal-poor GCs are
universally old, the metal-rich GCs in ellipticals can have a large
spread in age (Puzia 2002; Larsen \etal 2003).  The story told by the few GC systems with
substantial kinematic data is less consistent.  In general, the
metal-poor GC sub-populations have a higher velocity dispersion, but
rotation in either
population varies on a galaxy-by-galaxy basis (Zepf \etal 2000,
C\^{o}t\'{e} \etal 2001, Perrett \etal 2002, C\^{o}t\'{e} \etal 2003).

Globular clusters are in many ways ideal test particles for probing the
dynamics of early-type galaxies.  While integrated long slit
spectroscopy of the galaxy light is limited to $r<2r_e$ due to rapidly
declining surface brightness, GCs are essentially unresolved and are
easily observed at large distances from the galaxy center.
As single-age, single-metallicity entities, GCs also offer the
opportunity to correlate age and chemical enrichment with dynamics,
providing a powerful probe of galaxy evolution.  Moreover, luminous
ellipticals can host thousands of GCs, providing ample statistics.
Although GCs are often considered ideal tracers of ancient 
star formation because they are simple stellar populations, they still
only make up less than 1\% of the light in a typical galaxy.
Comparisons to the field star population are essential for any complete
picture of galaxy formation.  

One of the best galaxies for conducting an in-depth study is the
nearby post-merger elliptical \cena\ (Centaurus~A).  At a distance of
3.5~Mpc (Hui \etal 1993), \cena\ is the nearest giant elliptical
galaxy, and provides a unique opportunity to investigate both the nature of
spheroids and the effects of the merging process.  
In ellipticals, only for 
\cena\ has the metallicity distribution of the {\it field} population
been determined (W.Harris \& G.Harris 2002).  Likewise, only a handful
of galaxies have both globular cluster and planetary nebula (PN)
kinematic data (Romanowsky \etal 2002). 
In this paper, we present results based on the GC survey described in 
Paper I (Peng, Ford, \& Freeman 2004b).  We combine these results
with those presented in our survey of planetary nebulae (Peng, Ford, \&
Freeman 2004a) to try and elucidate the formation history of this
interesting and important galaxy.

\section{The Colors of \cena\ GCs}

The optical colors of old GCs are often used as proxies for
metallicity.  This is because for GCs of similar age, the main property
that drives the shape of the spectral energy distribution is metal
content.  Structure in the GC color distribution is thus taken as a
signature of various ancient epochs of metal-enrichment and star
formation.

We present multiple color distributions for the \cena\ GCs using our
broad wavelength coverage.  All GC colors have been
dereddened using the reddening maps of Schlegel, Finkbeiner, \& Davis
(1998).  Using code made publicly available by David Schlegel, we
determined the foreground reddening toward each individual GC.
Although the mean reddening is $E(B-V) = 0.115$, 
because we are concerned with the $U$-band,
the variations in extinction across two degrees of sky
can be significant --- they are at the level of 0.1~mag (peak-to-peak) in
$U$--$V$.  Jablonka \etal (1996) independently measured 
extinction variations in \cena\
that they attributed mainly to the foreground.  
We made no effort to correct for extinction
internal to \cena.  Therefore, it is
possible that some of the GCs are affected by additional
extinction, especially in the inner regions where there is clearly dust.
All the results in the following section must be kept with this caveat
in mind.

\begin{figure}
\plotone{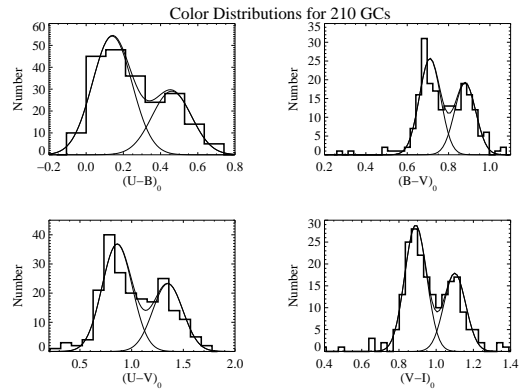}
\caption[Color histograms of spectroscopically confirmed GCs]
{\small Color histograms of spectroscopically confirmed GCs with Mosaic
photometry.  Colors presented are (\ub)$_0$,
(\bv)$_0$, (\uv)$_0$, and (\vi)$_0$.  Bin sizes were chosen to match the
median photometric error in each color.  All color distributions, 
exhibit evidence for bimodality.  Curves represent normally distributed
blue and red populations, and their sum, as determined by the KMM test. 
\label{figure:gchists_color}}
\end{figure}

We show the various color histograms for the \cena\
GCs. Figure~\ref{figure:gchists_color} plots four colors that are often
used --- (\ub)$_0$, (\bv)$_0$, (\uv)$_0$, and (\vi)$_0$.  To suppress
spurious structure in the histograms, the bin size for each color
is the median photometric error.  In theory, (\vi) is not necessarily
the best color for stellar population studies, but it has traditionally
had the smallest errors as a fraction of the spanned color range 
for all the optical colors.  

All color distributions show clear bimodal structure.  This is similar
to what was seen previously by Harris \etal (1992, hereafter HGHH92)
and Rejkuba (2001), and which has been seen in many other GC systems
(e.g.\ Larsen \etal 2001; Kundu \& Whitmore 2001).  
We use the KMM test (Ashman, Bird, \& Zepf
1994) to quantify the degree of bimodality.  The KMM test determines the
likelihood that a sample is better represented by two (or more)
Gaussians rather than by one.  We consider only the homoscedastic case
(both Gaussians have the same width), as the heteroscedastic case
is less well constrained.  We also clip the outlier GCs because they
adversely skew the distribution --- the extreme blue GCs are likely
young, and the extreme red GCs may suffer from internal reddening.
Using KMM, we find the GC distribution in 
all colors is very likely bimodal, with $P$ values near zero ($<0.001$).
The individual and summed sub-distributions can be seen as curves
plotted over the histograms in Figure~\ref{figure:gchists_color}.  For
the colors (\ub)$_0$, (\bv)$_0$, (\uv)$_0$, and (\vi)$_0$, the
blue peaks are determined to be at 0.14, 0.71, 0.86, 0.89 mag, and the red
peaks are at 0.46, 0.88, 1.35, 1.10 mag, respectively. 
The widths of the peaks, $\sigma=0.11,
0.05, 0.15, 0.06$~mag respectively, are close to the values expected if
they were dominated by photometric error.

We divided the sample into blue and
red GCs based on various criteria, and finally decided on a 
cut at (\vi)$_0 = 1.0$.  The extreme blue outliers in (\bv) and (\vi)
are likely to be young clusters and not extremely metal-poor.  This is
supported by the fact that the $U$ colors, which are not as sensitive to
age, do not show such extreme blue colors for these objects.

\section{The GC Radial Distributions}

\begin{figure}
\plotone{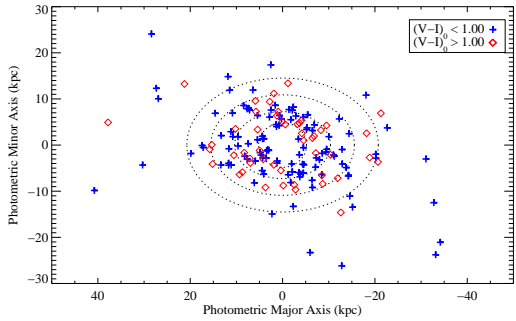}
\caption[Spatial distribution of all spectroscopically confirmed GCs]
{\small Spatial distribution of all spectroscopically confirmed GCs.
The coordinates are in kiloparsecs along the major and minor axes with
the respect to the galaxy center.  The ellipses represent approximate
isophotes for the galaxy at 1--4~$r_e$.  The GCs are coded by their
(\vi)$_0$ color, divided into two populations.
\label{figure:spacegc_spatial}}
\end{figure}

\begin{figure}
\plotone{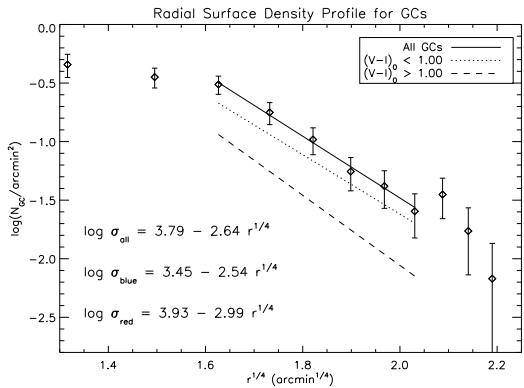}
\caption[Radial surface density distributions of known GCs]
{\small Simple radial distributions of known GCs.  Surface density of known
GCs in circular annuli of 1\arcmin\ thickness is plotted versus $r^{1/4}$.  
A de Vaucouleurs profile provides
a fairly good description of the radial profile, and the solid line
provides the best fit to the points least affected by incompleteness
($6\arcmin < r_{proj} < 18\arcmin$).  
The profiles for the red and blue GCs are also overplotted, and the
best fit linear coefficients are shown.
\label{figure:spacegc_radial}}
\end{figure}

Figure~\ref{figure:spacegc_spatial} plots the locations of each
confirmed GC with respect to the major and minor axes.  Another 
representation of the locations of the GCs are plotted in 
Peng \etal (2004b), Figure~10.
The symbols are coded to represent the blue and red GC populations.
Figure~\ref{figure:spacegc_radial} shows the approximate radial surface
density profiles
for the GCs and their blue and red subcomponents.  The surface density
was computed in circular annular bins, and fit with a de
Vaucouleurs profile.  While the galaxy is obviously non-circular,
especially in the halo, the
ellipticity of the GC system is not well-determined, and so we have
assumed the simplest case.

We do not correct for incompleteness or selection
biases.  Because of our requirement that GCs must be spectroscopically
confirmed, the selection function for our sample is quite complicated.
However, to first order, we assume that these biases do not vary
strongly as a function of radius except for in the very inner regions.  
Incompleteness is a problem in
the center due to the complex structure of the dust lane and the
bright background from the galaxy.  These bins show a drop-off in
surface density (where we would expect it be rising), and are ignored for the
purposes of fitting the profile.

\begin{figure}
\plotone{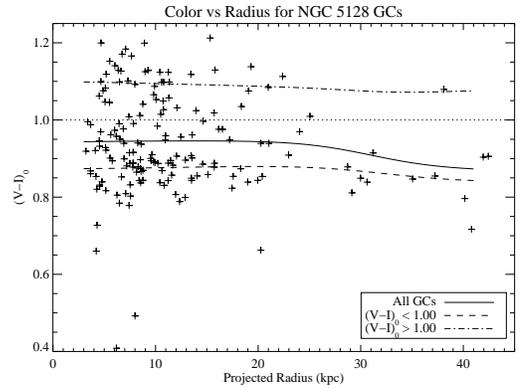}
\caption[(\vi)$_0$ versus Projected Radius]
{(\vi)$_0$ as a function of projected radius.  The red and blue
populations are divided at (\vi)$_0 = 1.00$, which is shown by the
horizontal dashed line.  The solid line is the kernel smoothed average
color as a function of radius.  The dot-dashed and dashed lines 
represent the same
for the red and blue populations.  All smoothing was done with a 14~kpc
wide gaussian kernel.  Most of the GCs beyond 25~kpc are blue, creating
an overall negative color gradient in the GCs.  Each individual
population, however, shows little evidence for a color gradient.
\label{figure:spacegc_colrad}}
\end{figure}

The profile of the entire GC population is well fit by a de Vaucouleurs
law.  Also overplotted are the fitted profiles to the red and blue
sub-populations.  The slope of the red population is formally
steeper (more centrally concentrated) than that of the blue population.
However, a more detailed treatment may
confirm or reject this spatial dichotomy,
considering our simplistic assumptions.  
Investigations in other ellipticals (e.g.\ Geisler, Lee, \& Kim 1996; 
Neilsen, Tsvetanov, \& Ford 1997)
show that the red populations are typically more centrally
concentrated that the blue, and that these dual spatial scales are the
driving reason behind metallicity gradients seen in GC systems as a
whole.  We show this more clearly in Figure~\ref{figure:spacegc_colrad} where
we plot the (\vi)$_0$ color versus the projected radius.  Nearly all of
the GCs beyond 25~kpc are in the blue sub-population.  This is partially
because there are more blue GCs in general, and so they are
statistically more likely to be present in the outer regions.  The three lines
represent the kernel smoothed (\vi)$_0$ color as a function of radius.
There is a slight color gradient in the overall population, but the blue
and red populations individually show flat color profiles.  This is
similar to the case of Galactic GCs, for which there is no
significant gradient in metallicity for either bulge/disk or halo clusters.

\section{Relative Metallicities and Ages}
Massive star clusters form during episodes of moderate to intense
star formation in a galaxy (e.g.\ Larsen \& Richtler 1999).  Because
larger numbers of massive clusters form in more massive starbursts
(Whitmore 2000), the population of surviving clusters which
eventually become GCs should trace their host galaxy's major epochs of
star formation.  As discrete, identifiable collections of stars with
a single age and metallicity, these clusters are also the simplest stellar
populations to model.  Together, these properties offer the hope that
the study of GC systems will ultimately help reveal the star formation
and metal enrichment history of galaxies.  In this section, we attempt
to derive physical quantities from the observables by comparison to
evolutionary synthesis models.

\subsection{Photometry} \label{section:agez_phot}
Despite the relative simplicity of GCs with respect to the integrated
stellar light of an entire galaxy, disentangling the effects of age and
metallicity on observed properties is still difficult.  Modelers can
calculate the photometric and spectroscopic evolution of simple stellar
populations as a function of age for different metallicities
(e.g.\ Bruzual \& Charlot 1993, Vazdekis 1999, 
Yi \etal\ 2001).  Examples of these
age-metallicity grids in color-color space are taken from the 2001
release of the Bruzual \& Charlot models (BC01).
Unfortunately, there is a severe and well-known degeneracy between age
and metallicity effects in $BVRI$ colors for stellar populations older
than 1--2~Gyr.  However, as Yi (2003) has pointed out, the addition of
information in the $U$-band makes optical colors somewhat less prone to
this degeneracy, at least at low metallicities.

\begin{figure}
\plotone{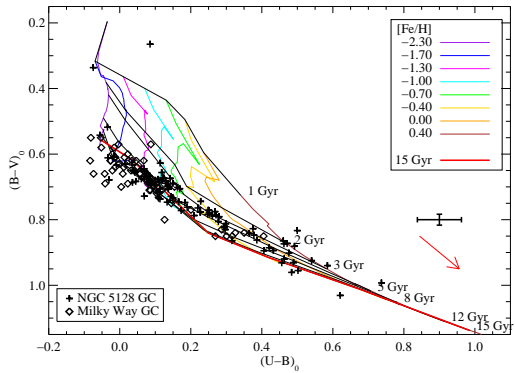}
\caption[NGC~5128 GCs and Milky Way GCs on Bruzual-Charlot (version 2001)
Age-Metallicity Grids]
{\small NGC~5128 GCs and Milky Way GCs on Bruzual-Charlot (version 2001)
Age-Metallicity Grids.  The grids are oriented such that, roughly, 
metallicity increases towards the right, and age increases towards the
bottom. Cross points are 105 \cena\ GCs with  
$U_{err} < 0.1$~mag, and diamond points
are Galactic GCs that have low reddening ($E(B-V) < 0.3$).  The two
samples match at low metallicities, showing that most metal-poor GCs in
\cena\ are likely as old as the Galactic GCs.  At higher metallicities,
the age and metallicity are still degenerate, even with $U$-band data,
and errors in both photometry and reddening make age determinations or
comparisons difficult.
Also displayed in the lower right are median error bars and a
reddening arrow equivalent to $E(B-V) = 0.1$.  The two points in the
upper left are the young clusters HGHH-G279 and pff\_gc-029, which are
described in the text.
\label{figure:gc_bcgrids_ubv}}
\end{figure}

The utility of $U$ observations is offset by the difficulty in obtaining 
deep and accurate $U$-band photometry.
Atmospheric extinction and the rapid loss of sensitivity
of current astronomical detectors at shorter wavelengths makes
precise $U-$band measurements notoriously difficult.  Much of the
ground-breaking work on GC systems in the 1990s was done with $HST$ on
elliptical galaxies at distances greater than 10~Mpc.  Unfortunately,
the relatively low response of the WFPC2 detectors in the blue and
near-ultraviolet required most $HST$ work to be done in the $V$ and
$I$ (or equivalent) bandpasses.  Moreover, these galaxies are too
distant for ground-based 4-meter class telescopes.  Recently, though, 
Rejkuba (2001) obtained $U$ and $V$ photometry of two fields in \cena\
using the 8-meter VLT and used the $U$-band's sensitivity to
metallicity.  Jord{\'a}n \etal (2002) obtained Str\"{o}mgren {\it u}
photometry of M87 GCs and used it in conjunction with other filters to
constrain the relative ages of the metal-rich and metal-poor GC
populations.  Whitlock, Forbes, \& Beasley (2003) also took $U$
observations of GCs in NGC~3379, comparing $UBRI$ colors to model grids. 
However, $U$ photometry is still relatively rare and the
large number of confirmed \cena\ GCs makes our data set interesting.

For the purposes of this paper, we simply compare the $UBV$ colors of
the \cena\ GCs to Milky Way clusters, and to the BC01 model grids in
order to explore global trends in age and metallicity.
We defer a detailed discussion of the use of $UBV$ colors to
derive ages and metallicities to another paper (Yi \etal 2003).
In Figure~\ref{figure:gc_bcgrids_ubv}, we show the $UBV$ colors of 105
\cena\ GCs with respect to the dereddened colors of the Milky Way GCs 
(W.Harris 1996) and the BC01 grid (we have also compared to the UBV
grids of Maraston (2003, in prep), but the differences between the two
sets of models are not important for our purposes).  Those
plotted are \cena\ GCs with $U_{err} < 0.1$~mag and Galactic GCs with
$E(B-V) < 0.3$.  A caveat is that both of these cuts are slightly biased
against the most metal-rich GCs.  In \cena, these clusters have the reddest
{\ub} colors and so will have fainter $U$ magnitudes for a fixed $V$.  
In the Galaxy, the more metal-rich clusters are preferentially toward the
bulge, and so will have higher reddening.  We know from 
color-magnitude diagrams (CMDs) of Galactic GCs that these clusters have a
large range in metallicity but are all older than $\sim10$~Gyr, a result
generally reproduced by the model grids.

Figure~\ref{figure:gc_bcgrids_ubv} shows that our photometry is precise enough
to make comparisons to both the models and the Galactic GCs at the
metal-poor end.  The metal-poor clusters in \cena, 
toward the left of the diagram,
are consistent with being as old as the metal-poor Milky Way GCs.
These GCs also happen to fall on the 12--15~Gyr tracks in the BC01 models.
Unfortunately, at low metallicity this age is degenerate with younger
ages because of the contribution of blue horizontal branch stars to the
broadband colors.  However, we know independently that the metal-poor
Galactic GCs with these colors are closer to 15~Gyr in age than to 5~Gyr
so it is likely that \cena\ clusters with these colors are also old.
These GCs are well-separated from two
clusters that are much younger than the rest, with ages~$\leq 1$~Gyr.
These younger clusters are HGHH-G279, with (\bv)$_0 = 0.34$, and pff\_gc-029,
with (\bv)$_0 = 0.26$.
Held \etal (1997) spectroscopically determined that HGHH-G279 may be
young.  Also, the youngest of these star clusters (pff\_gc-029) belongs
to the young blue tidal stream described in Peng, Ford, Freeman \&
White (2002).

At the metal-rich end, however, we are handicapped by the increasing
degeneracy between age and metallicity, and also by the lack of Galactic
GCs with low reddening.  According to the grids alone, the locus of 
\cena\ GCs appear to be above the 12 and 15~Gyr isochrones, but the few
Galactic GCs in this range appear to occupy a similar locus.  Given the
errors in photometry and reddening, we conclude that our photometry has
only weak leverage on the ages of the metal-rich GCs.  For a more
definitive answer, we require spectroscopy.

\subsection{Spectral Line Indices} \label{section:agez_spec}

While the signal-to-noise for broadband photometry is much higher than
for spectroscopy, the low spectral resolution limits astronomers to
studying trends in the shape of the spectral energy distribution.
Spectroscopy is a complementary approach that permits the measurement of
individual spectral features that are sensitive to temperature and
elemental abundances.  Because each cluster in our sample was required 
to have a measured radial velocity to be included in our
catalog, we are fortunate to have spectra for most \cena\ GCs.  For the
brighter GCs, the measurement of line indices provides a test independent
of our photometry for the relative ages of the GC sub-populations.

We use the Lick/IDS system of optical spectral line indices defined by
Worthey \etal (1994) and Worthy \& Ottaviani (1997).  These indices
cover many important metal and Balmer lines in the blue half of the
optical spectrum.  These authors and others have combined the original
Lick/IDS stellar spectra using models for stellar evolution to produce
model spectra of integrated stellar populations.  By comparing index
measurements in our spectra to the model grids for an
instantaneous burst of star formation at different ages and
metallicities, we hope to constrain these quantities in the \cena\ GCs.

\subsubsection{Transforming to the Lick/IDS System}

Before making this comparison, it is important to prepare the spectra
properly and place the measurements on the Lick system.  All GC spectra
had any underlying galaxy light subtracted (see discussion in 
Peng \etal 2004b).  The median level of galaxy light contamination
is 11\%, and this subtraction is only important for a few extreme cases.  
For the purpose of matching the Lick/IDS system, we observed five
standard stars HD140283, HR5196, HD64606, HR5568, HR2574.  
These stars span a range of metallicity $-2.45 < {\rm [Fe/H]} < +0.03$
and effective temperature $4019 < T_{eff} < 5650$.
HD140283 was observed twice, each time through a different fiber.
We de-redshifted these stars to the rest-frame using velocities listed
in SIMBAD.  The original Lick/IDS stars were observed at a resolution of 
8--11~\AA\ between 4000-6000\AA, so we smoothed our spectra with a
variable-width Gaussian kernel to match the resolution given in
Worthey \& Ottaviani (1997).

\begin{figure}
\plotone{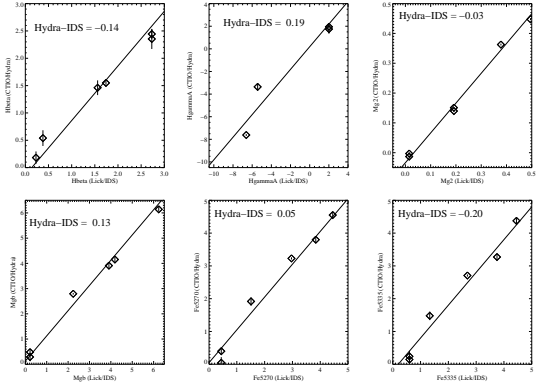}
\caption[Offsets between CTIO/Hydra and Lick/IDS system]
{\small Determining Index Offsets between CTIO/Hydra and Lick/IDS using
standard stars HD140283 and HR5196.  The solid line is constrained to
have a slope of 1.
\label{figure:lickoffsets}}
\end{figure}

By comparing our index measurements for these stars with the
measurements of Worthey \etal, we derived offsets for each line index
that could then be applied to our CTIO/Hydra measurements.  This
comparison is shown for six indices in Figure~\ref{figure:lickoffsets}.

\subsubsection{Line Index Measurements: Age and Metallicity}

For each GC spectrum, we measured all of the line indices from Worthey
\etal (1994) and Worthey \& Ottaviani (1997).  We confine our discussion
to a few of the indices that are most useful.  As has been discussed
extensively in the literature by these and other authors, 
the age-metallicity degeneracy
also plagues optical spectroscopic work.  However, it is possible to
somewhat disentangle these two, albeit to a relatively low precision,
using lines that are more sensitive to either age or metallicity.  We
use the \mgfe\ index defined by Thomas, Maraston, \& Bender 
(2003; hereafter TMB03) as our
primary metallicity indicator.  This index is a modified version of the
one described by Gonz\'{a}lez (1993) and is defined as 
$[MgFe]^{\prime} = 
\sqrt{Mgb\times(0.72\times {\rm Fe5270} + 0.28\times {\rm Fe5335}) }$.
By combining an alpha-element (Mg)
with iron, this index is insensitive to the effects of
alpha-enhancement when using the TMB03 models.
As our primary age index, we use \Hb, the Balmer line with the
highest signal-to-noise, although we also show results for the \Hga\
index, which some have argued may be a better age indicator
(e.g.\ Vazdekis \& Arimoto 1999, Puzia \etal 2002).

\begin{figure}
\plotone{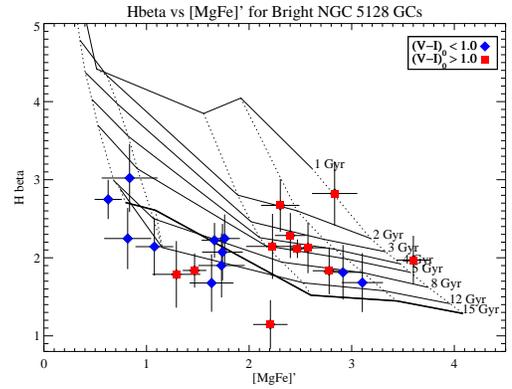}
\caption[\Hb\ versus MgFe: Bright GCs]
{\small \Hb\ and \mgfe\ Lick/IDS index measurements 
overplotted on the age
and metallicity model grids from TMB03.  Dotted lines (vertical) represent
constant metallicity with values [m/H] = [-2.25, -1.35, -0.33, 0.0, +0.35]
(left to right).  The grid is for [$\alpha/$Fe] = 0.
Solid lines (horizontal) represent constant age with
values from 1--15~Gyr as labeled on the right side.  The points are
coded (blue diamonds and red squares) to represent the (\vi)$_0$ color
sub-population to which each GC belongs. This figure shows the index 
measurements for 23 bright, (S/N)$_{{\rm H}\beta} >40$, GCs.   
The metal-poor GCs are universally old, but the metal-rich GCs have a
younger range of ages.
\label{figure:hb_mgfe_bright}}
\end{figure}

\begin{figure}
\plotone{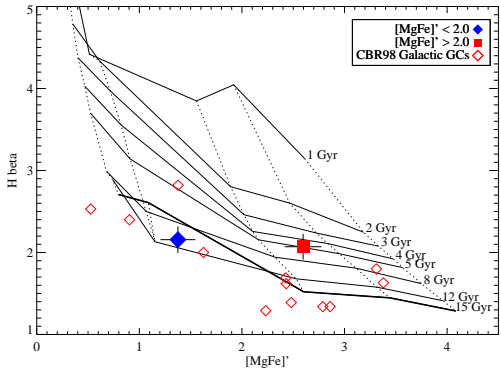}
\caption[\Hb\ versus MgFe: Coadded Spectra]
{\small \Hb\ and \mgfe\ Lick/IDS index measurements overplotted 
on the age and metallicity model grids from TMB03.  The grid is the 
same as for Figure~\ref{figure:hb_mgfe_bright}. 
The figure shows the index measurements for two GC spectral
templates created by coadding GCs in the same \mgfe\ bin.  Also
overplotted with open symbols in this figure are index measurements for
Galactic GCs from CBR98.
\label{figure:hb_mgfe_coadd}}
\end{figure}

In Figures~\ref{figure:hb_mgfe_bright} and \ref{figure:hb_mgfe_coadd}, 
we compare our measurements with the model grids of TMB03 and in the
latter, to the Galactic GC observations of Cohen, 
Blakeslee, \& Ryzhov (1998, CBR98).  For the TMB03 grids, we choose the
models with canonical mass-loss on the RGB (blue horizontal branches).
When comparing this model grid with the
BC01 grid in Figure~\ref{figure:gc_bcgrids_ubv}, one can see how the 
effects of age and metallicity, while still covariant, are much less
entangled when using spectral index measurements than they are with
optical broadband colors.  

Figure~\ref{figure:hb_mgfe_bright} shows \Hb\ and \mgfe\ index
measurements for all GCs with CTIO/Hydra spectra that have a
signal-to-noise (S/N) $> 40$ per resolution element in the Lick \Hb\
bandpass.  This roughly corresponds to errors of 
$\sigma_{{\rm H}\beta} < 0.4$,
and $V<18.8$.  Errors were determined using counting
statistics and read noise in the original spectra.  They do not
incorporate systematic errors that may be inherent to our transformation
to the Lick/IDS system.  

In this figure, the points are coded by
their (\vi)$_0$ color in order to provide a visual aid for
comparison between our photometry and spectroscopy.  
As expected, the broadband color correlates with
metallicity --- while mixing between the two groups is unavoidable given
the overlapping nature of the two distributions, 
the blue GCs are generally metal-poor and the red GCs are
metal-rich.  The metal-poor clusters are
consistent with being a universally old population (8--15~Gyr old).
The metal-rich GCs, however, appear to have a large spread in ages from
1--10~Gyr, with the bulk of them being significantly
young.  In fact, of the metal-rich GCs with line index
measurements, only one-third have derived ages older than 8~Gyr.  

\begin{figure}
\plotone{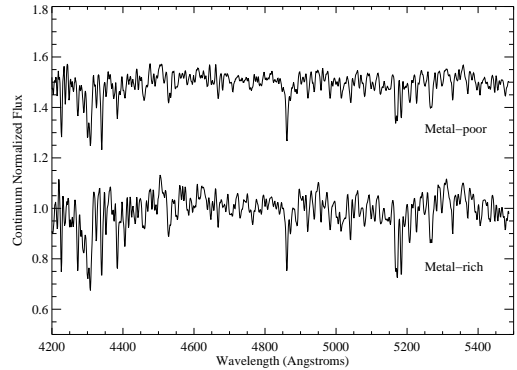}
\caption[Composite GC Spectra]
{\small Two composite GC spectra, created from bins of \mgfe.  Each
composite is the sum of the spectra of $\sim$10 bright GCs.  The counts on
the y-axis reflect continuum-normalized flux.  In the case of the upper
(metal-poor) spectrum, 0.5 has been added to visually separate it from the
metal-rich composite.  
The line indices do show some of the expected trends.  In particular, the Mg
region around 5200\AA\ and the nearby Fe lines are stronger for redder
GCs.  These two spectra are the composites for which Lick index
measurements were made and are plotted in
Figure~\ref{figure:hb_mgfe_coadd}.
\label{figure:gccomp}}
\end{figure}

To determine the mean ages of these populations, we created two composite
spectra in bins of \mgfe, dividing at a value of 2.  This is
slightly more precise than dividing by (\vi)$_0$, achieving the same effect.
Each one of these composites is the sum of all the high S/N spectra in that
metallicity bin.  The composite spectra are plotted in
Figure~\ref{figure:gccomp}.  The metal lines, especially Mg and
Fe are stronger in the metal-rich composite, but \Hb\ is of
comparable strength in both spectra.

\begin{figure}
\plotone{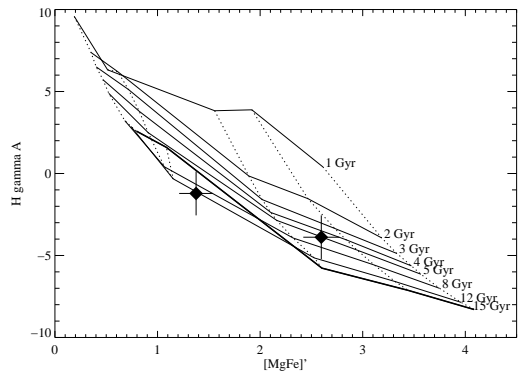}
\caption[\Hga\ versus MgFe: Coadded Spectra]
{\small \Hga\ and \mgfe\ Lick/IDS index measurements overplotted 
on the age and metallicity model grids from TMB03.  The grid is from the
TMB03 models, and is essentially the
same as for Figure~\ref{figure:hb_mgfe_coadd}. 
The figure shows the index measurements for two GC spectral
templates created by coadding GCs in the same \mgfe\ bin.
The mean ages determined from \Hga\ agree with the ages derived from \Hb.
\label{figure:hga_mgfe_coadd}}
\end{figure}

We measured the same indices for these composites and plot their values
in Figure~\ref{figure:hb_mgfe_coadd}.  The 1-sigma errors here 
are a combination
of both random errors and systematic errors, the latter of which are
from the transformation to the Lick/IDS system.  The random errors, as
determined from the bootstrap, are comparable to the size of the data points.
The systematic errors are dominant, and were determined by the scatter
about the mean offset value for the observed standard stars.
Also overplotted in this panel are the Lick
index values for a few Galactic GCs as measured by CBR98.  
Again, while the metal-poor GCs in \cena\ are
consistent with the old ages of the Galactic GCs, the more metal-rich
\cena\ GCs are younger than their Galactic counterparts.  We see the
same trend when we measure the H$\gamma_A$ index, both on individual
spectra and the composites.  This can be seen in
Figure~\ref{figure:hga_mgfe_coadd}.  

There is an age degeneracy at old ages for metal-poor GCs due to the
contribution of flux from blue horizontal branch stars.  Whichever way
we break this degeneracy, the metal-poor clusters are old, with ages of
10 or 13~Gyr with an error of $\pm2$~Gyr.  The metal-poor composite also 
has [m/H]$\sim -1.1$. The metal-rich composite has an age 
of $5^{+3}_{-2}$~Gyr, with [m/H]$\sim -0.1$.  Regardless of the age
degeneracy at low metallicities, the metal-rich clusters appear
significantly younger (by 5--8~Gyr) than their metal-poor counterparts.  

One caveat that must be mentioned is that old GCs with
prominent blue horizontal branches (BHBs) can have spectra that mimic
those of intermediate-age stellar populations.
For metal-poor GCs, this causes a degeneracy at old ages.  It is also
possible that strong Balmer line strengths in metal-rich GCs
can be explained by a significant number of BHB stars (e.g.\ Peterson
\etal 2003).  Both de Freitas Pacheco \& Barbuy (1995) 
and Beasley, Hoyle, \& Sharples (2002) show that blue horizontal
branches can enhance \Hb\ line strengths by up to 1.0\AA.  
However, it is unlikely that all of the metal-rich GCs are
``second parameter'' clusters, although it would certainly be
interesting if that were the case.  
Moreover, a few of the very youngest clusters, such as the one discussed
in Peng \etal (2002), have \Hb\ line strengths that are too high to be fully
explained by a BHB contribution to a $10+$~Gyr old population.  

As with all comparisons to model grids, these ages are to be trusted
more in a relative sense rather than an absolute sense.  Only the plots
involving composite spectra include the errors in the Lick/IDS stars, 
and none include errors in the model grids.
It is also possible that by limiting ourselves to bright GCs, we are
biasing ourselves in favor of younger clusters.
Despite these caveats, our spectroscopy shows that 
\Hb\ is systematically stronger in the metal-rich GCs.  
This is strongly
suggestive that there is an age difference between the metal-poor
and metal-rich GCs, and also that the metal-rich GCs are consistently
younger than the Galactic GCs.

If the metal-rich GCs are of intermediate age, then they should be more
luminous than if they were the same age as Galactic GCs.  Should we
expect to see exceptionally bright metal-rich GCs in \cena?
Metallicity, however, plays a competing role with age in
that a metal-rich cluster of the same mass as a
metal-poor counterpart will be intrinsically fainter.  The result is
that the BC01 models predict the turnover magnitude of the luminosity
function for an intermediate-age
metal-rich GC population with (\vi)$_0 = 1.1$ to be only
$\sim0.3$~mag  brighter in $V$ than for a metal-poor population with 
(\vi)$_0 = 0.89$.  The turnover magnitudes for the two populations in
$U$ should be nearly identical.  Unfortunately,
our spectroscopy does not go sufficiently past the turnover of the GC
luminosity function to determine the turnover magnitude for either
population.  However, within the sampled magnitude range, the
distribution of GC brightnesses is similar, which is consistent
with the age distribution we see.

\subsubsection{Line Index Measurements: Alpha-Element Enhancement}

The abundance ratio of alpha-elements to iron can shed light on the
duration of the star-forming event that created the observed stars.
This is because alpha-elements are preferentially created in Type~II
supernovae whose progenitors are massive stars.  Thus, alpha-elements
such as Mg and O primarily enrich the local interstellar medium (ISM) 
in the early stages of a star-forming event.  Type~I
supernovae, which preferentially produce iron, have low-mass progenitors
(white dwarf--red giant pairs) that take at least 1~Gyr to inject
their metals into the ISM.  The result is that stars created in
short bursts are likely to be enhanced in alpha-elements.  Stars that
formed in more prolonged star-forming events, or those that formed out
of gas pre-enriched with iron, will not have alpha-enhanced elemental
abundances.  Varying levels of alpha-enhancement can also mean that the
initial mass function of stars was different for different stellar
populations. 

\begin{figure}
\plotone{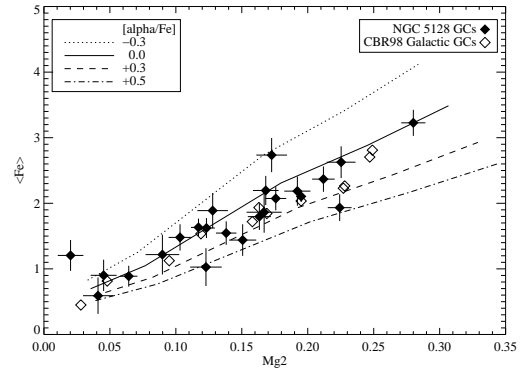}
\caption[$\langle{\rm Fe}\rangle$ versus Mg2: Bright GCs]
{\small $\langle{\rm Fe}\rangle$ and Mg2 Lick/IDS index measurements 
overplotted on the
$[\alpha/{\rm Fe}]$-metallicity model grids from TMB03.  The lines all
represent constant $[\alpha/{\rm Fe}]$ with the solid line being 
the solar value.  Like in Figure~\ref{figure:hb_mgfe_bright}, 
the metallicity along the lines ranges from [m/H]$ = -2.25$ to $+0.35$.  
The solid points with error bars are our measurements for \cena\ GCs.
The open symbols are the Galactic GCs from CBR98.
Plotted here are the index
measurements for 23 bright GCs that have (S/N)$_{{\rm H}\beta} > 40$.
\label{figure:fe_mg2_bright}}
\end{figure}

\begin{figure}
\plotone{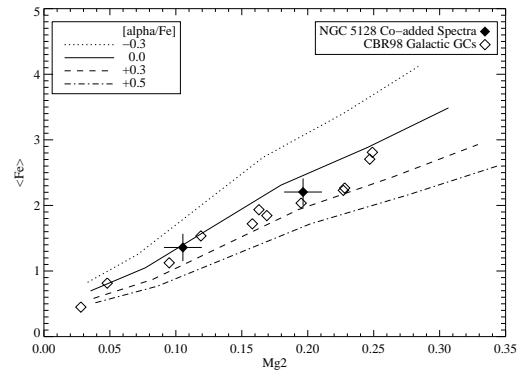}
\caption[$\langle{\rm Fe}\rangle$ versus Mg2: Composite GCs]
{\small $\langle{\rm Fe}\rangle$ and Mg2 Lick/IDS index measurements 
for composite GC spectra overplotted on the
$[\alpha/{\rm Fe}]$-metallicity model grids from TMB03.  Lines and
symbols are the same as for Figure~\ref{figure:fe_mg2_bright}.
\label{figure:fe_mg2_coadd}}
\end{figure}

\begin{figure}
\plotone{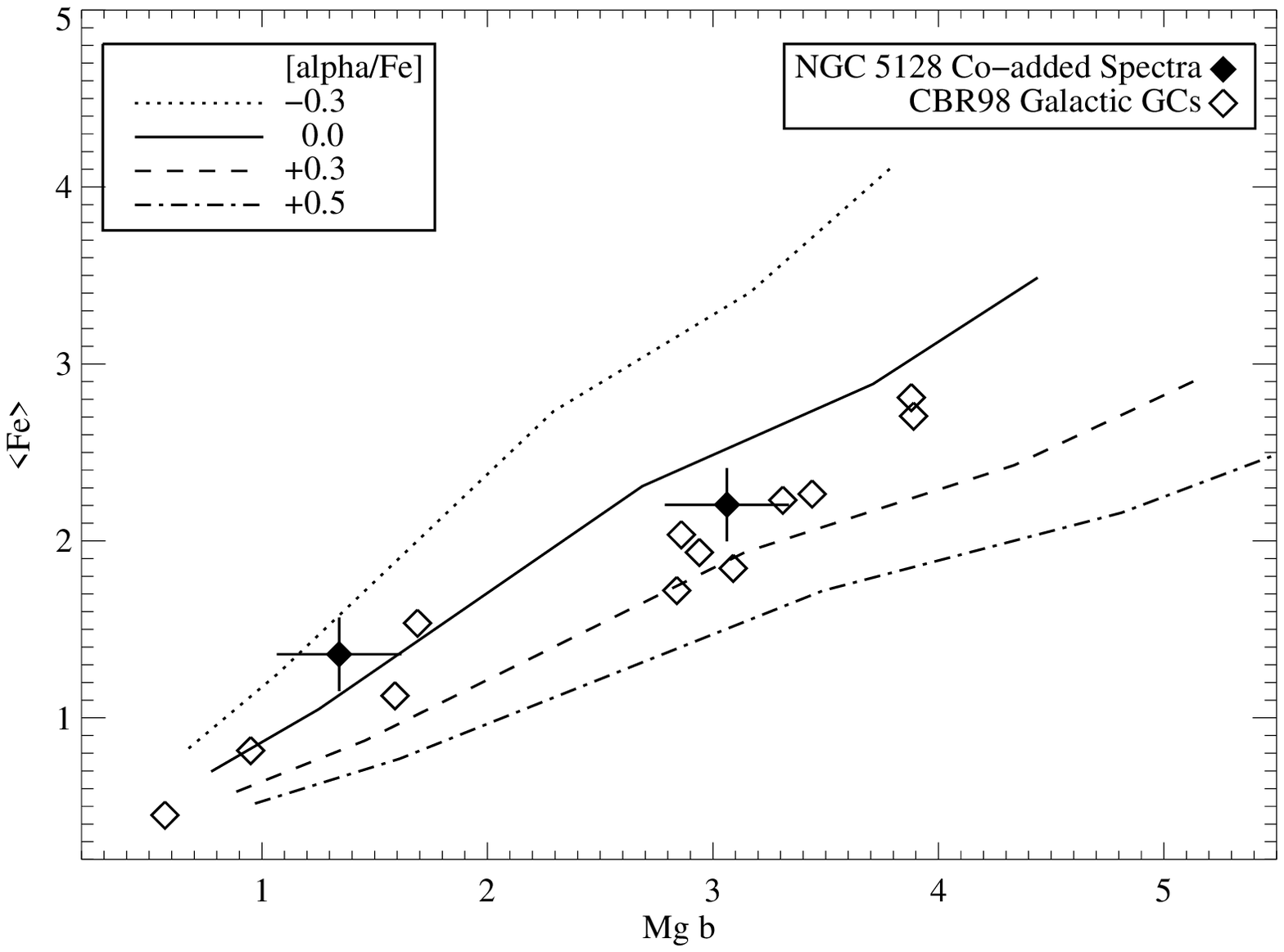}
\caption[$\langle{\rm Fe}\rangle$ versus Mgb: Composite GCs]
{\small $\langle{\rm Fe}\rangle$ and Mgb Lick/IDS index measurements 
for composite GC spectra overplotted on the
$[\alpha/{\rm Fe}]$-metallicity model grids from TMB03.  Lines and
symbols are the same as for Figure~\ref{figure:fe_mg2_bright}.
\label{figure:fe_mgb_coadd}}
\end{figure}

We attempt to constrain the level of alpha-element enhancement in the
\cena\ GCs.  TMB03 calculated Lick indices for different levels of
alpha-enhancement.  While these models are among the best available for
alpha/Fe analysis, one caveat is that they are poorly calibrated at
young and intermediate ages because they are matched to the Galactic
bulge GCs.  We use the Mg2 and combined $\langle$Fe$\rangle$ 
indices (where $\langle{\rm Fe}\rangle = ({\rm Fe5270}+{\rm Fe5335})/2$)
to compare alpha-element and iron abundances.
Figure~\ref{figure:fe_mg2_bright} shows the
TMB03 tracks for $[\alpha/{\rm Fe}] = $~[$-$0.3, 0.0, +0.3, +0.5] where 
$[\alpha/{\rm Fe}] = 0.0$ is
the solar ratio by definition and is represented by the solid line in
both panels.  Figure~\ref{figure:fe_mg2_bright} plots the bright (S/N$>40$)
\cena\ GCs and the CBR98 Galactic GCs on the TMB03 grid.  The Galactic
GCs are known (from spectroscopy of individual stars) to be generally 
enhanced in $\alpha$-elements (e.g.\ Habgood 2001) with 
$\langle[\alpha/$Fe$]\rangle = +0.3$.  This is 
shown in this comparison to the TMB03 models by their offset from
the solar value.  In the mean, all \cena\ GCs are consistent with having
the same $\alpha/$Fe enhancement as the Galactic GCs.  This is shown
more clearly in Figure~\ref{figure:fe_mg2_coadd}, where the indices have
been measure on the composite GC spectra.  As seen by the errors in this
figure, which include both random and systematic errors, this
conclusion is sensitive to the transformation used to place
our measurements on the Lick/IDS system.  As with the other composite
measurements, systematic errors dominate the random errors.  The same
trends are evident when the Mgb index is used to trace magnesium, as is
seen in Figure~\ref{figure:fe_mgb_coadd}.
The fact that these two indices give similar results is reassuring
because the Mgb index spans a smaller range of the spectrum than the Mg2
index, and is thus less likely to be affected by errors in the
transformation.  Going beyond measuring gross trends and
determining the {\it distribution} of $\alpha/$Fe values 
will require much higher quality data.

\section{Globular Cluster Kinematics}

The space motions of stars and star clusters in a galaxy provide another
window on the environment in which they formed.  Unlike gas, which
radiates energy in inelastic collisional encounters, stars and clusters
can be treated as collisionless particles.  Because of this, they retain for
a much longer period of time the space motions of the gas out of which
they formed.  This allows us to use their kinematics, not only as
tracers of the larger galactic gravitational potential, but as a probe
of their dynamical history.  In particular, the amount of angular
momentum in a given population may be an important clue to uncovering
the nature of its progenitors.
By adding a larger environmental context to the historical record,
kinematic data complements formation timescales and local
metal content that we have derived from their spectral energy distributions.

\cena\ is much too far away to measure proper motions of individual
objects, but we can measure the doppler line-of-sight velocity to obtain
kinematic information in one dimension.
Since our catalog is velocity-selected---we require that an
object have a radial velocity of $v_{helio}>250$~\kms\ in order to be
considered a GC---a kinematic analysis is a natural product of our
survey.  A comparison of GC kinematics to PN kinematics
has only been done for a handful of galaxies.  This is a
potentially valuable comparison as it can help determine whether the
GCs, or some subset of them, share a common formation history with
the bulk of the stars in the galaxy. 

\begin{figure}
\plotone{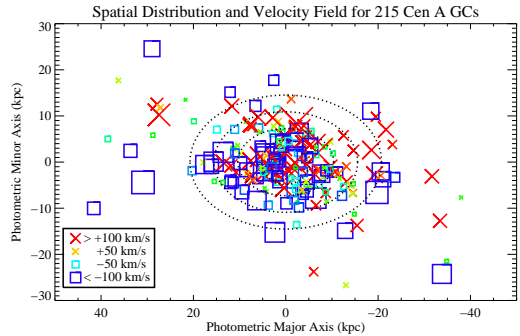}
\caption[Spatial Distribution and Velocities for 215 \cena\ GCs]
{\small Spatial distribution and velocities for 215 \cena\ GCs.  Dotted
ellipses represent the approximate isophotal contours of the stellar
light at 1--4~$r_e$.  Size and color represent the magnitude of the
velocity difference from systemic.  GCs with positive velocities have 
red 'x' symbols, while negative velocities have blue squares.  The
colors span a spectrum, with green points of either symbol representing
a velocity indistinguishable from systemic within the errors. This
figure shows the velocities of each individual GC.  
\label{figure:rvgc_all}}
\end{figure}

\begin{figure}
\plotone{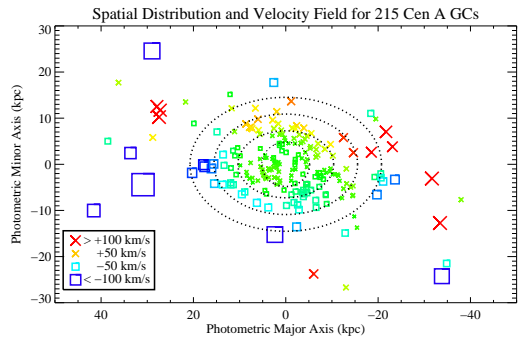}
\caption[Smoothed Velocity Field for 215 \cena\ GCs]
{\small Smoothed velocity field for 215 \cena\ GCs.  
Symbols and color coding are the same as for Figure~\ref{figure:rvgc_all}.
This figure plots the kernel smoothed velocity field.  Each symbol is at the
position of a known GC, but the velocity is now representative of the
smoothed field at that location with a 3~kpc gaussian kernel.
\label{figure:rvsmoothgc_all}}
\end{figure}

Figure~\ref{figure:rvgc_all} shows the GC velocity field.  The large
velocity dispersion of the GC system makes it hard to see any ordered
motion.  This is made more clear in Figure~\ref{figure:rvsmoothgc_all} where
the velocity field is smoothed with a Gaussian kernel ($\sigma=3$~kpc).
In this smoothed velocity field, the GC system shows a clear sign of
rotation outside of $\sim5$~kpc.  The rotation is around {\it both} the major
and minor axes, in a similar sense as we see for the PNe.

\begin{figure}
\epsscale{1.0}
\plotone{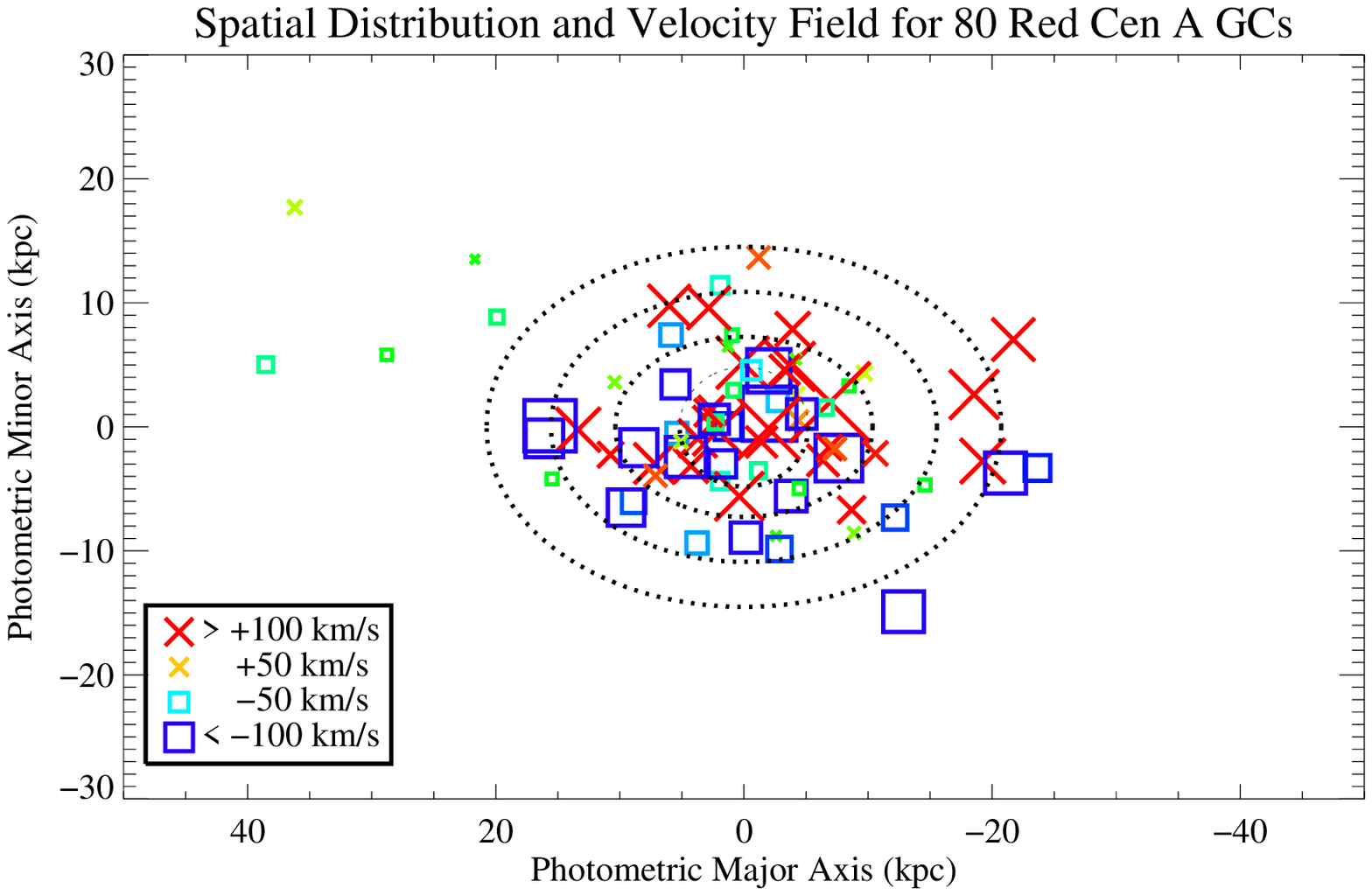}
\plotone{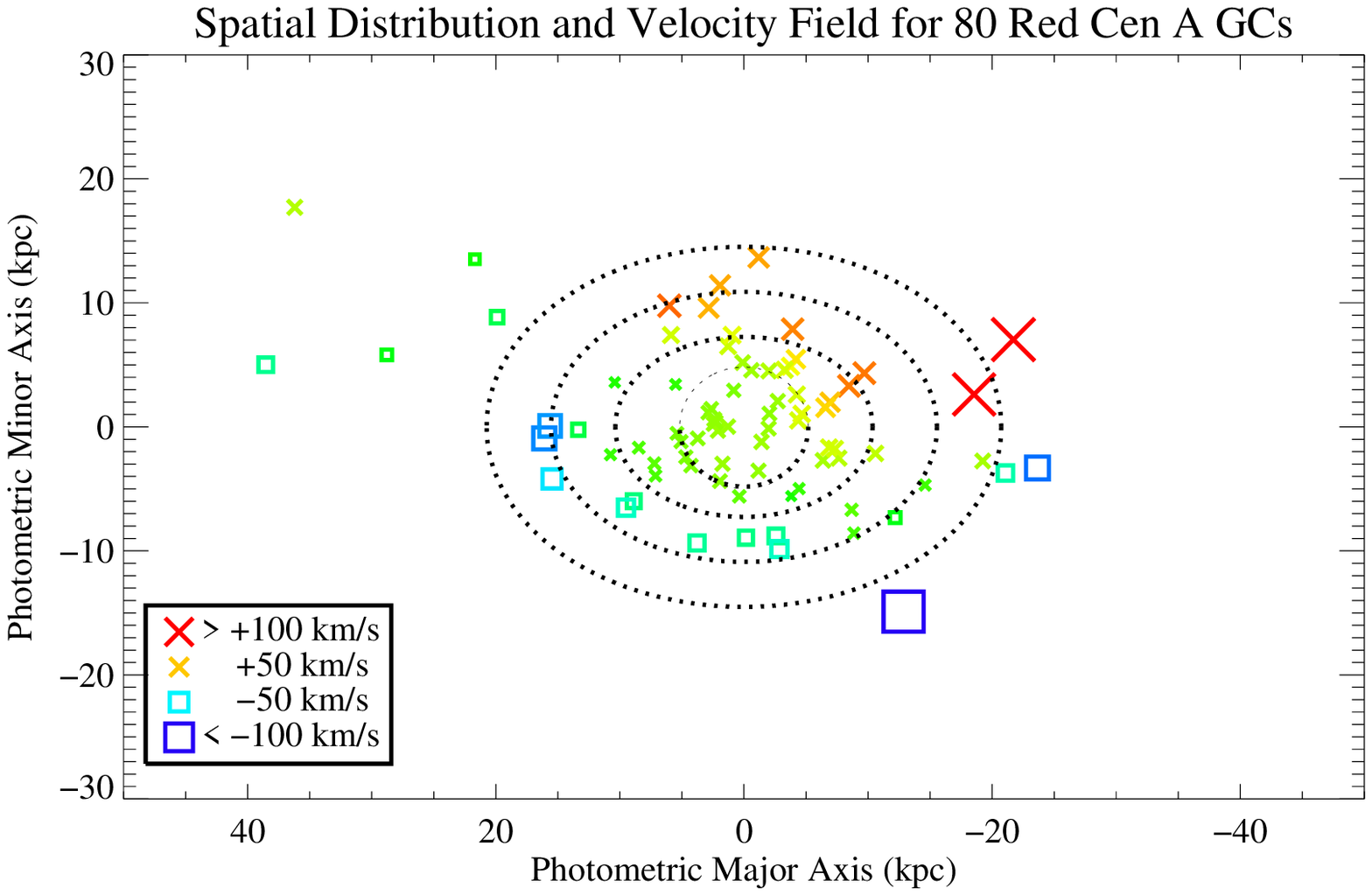}
\caption[Velocities for red (\vi)$_0 > 1.00$ GCs]
{\small Velocities for red (\vi)$_0 > 1.00$ GCs.  Most of these GCs
are within 15~kpc.  The top figure (a) shows the original measured
velocities and the bottom panel (b) shows the smoothed velocity field.
There is a clear rotation about both the major and minor axes.
\label{figure:rvsgc_red}}
\end{figure}

\begin{figure}
\plotone{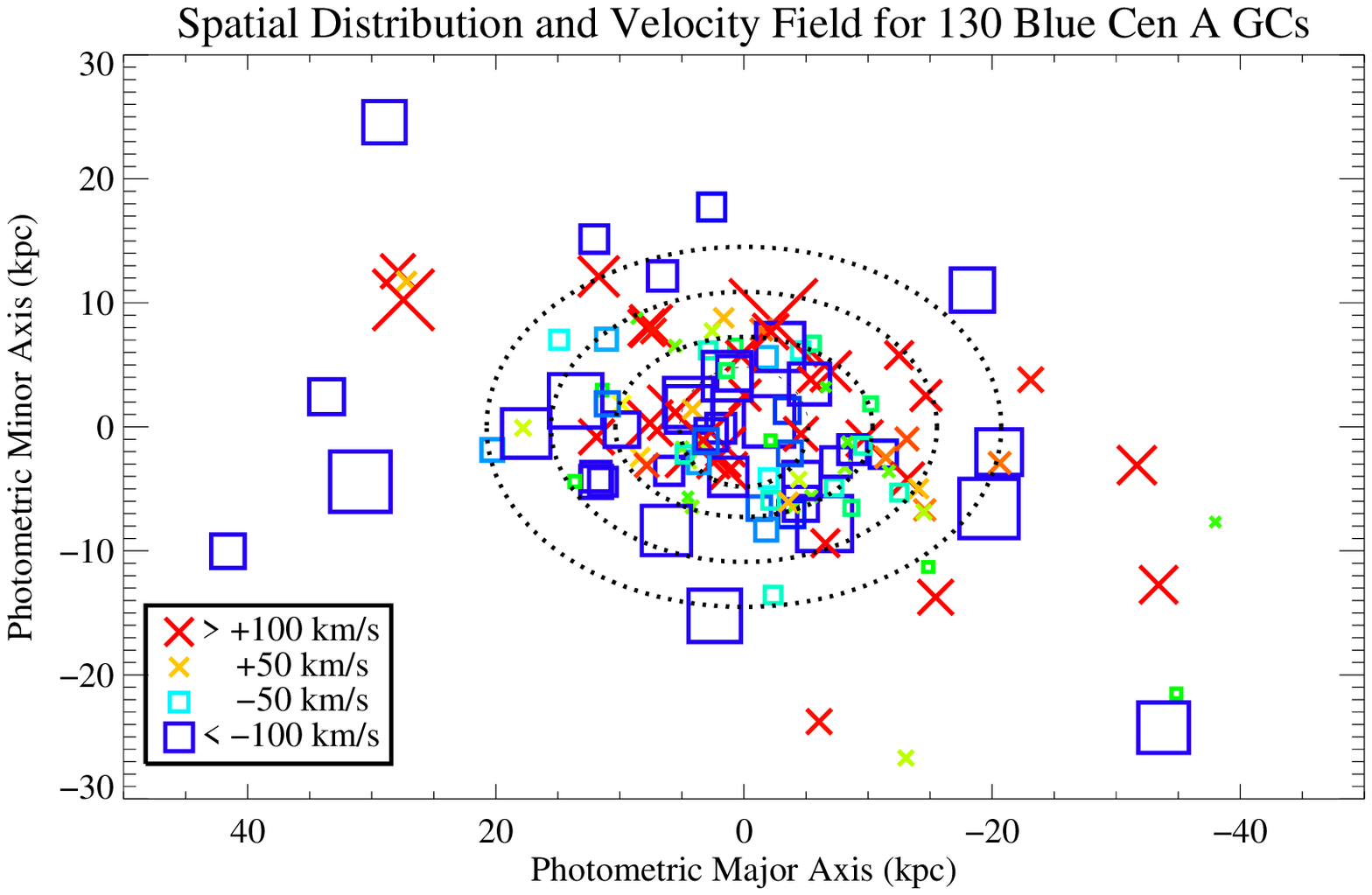}
\plotone{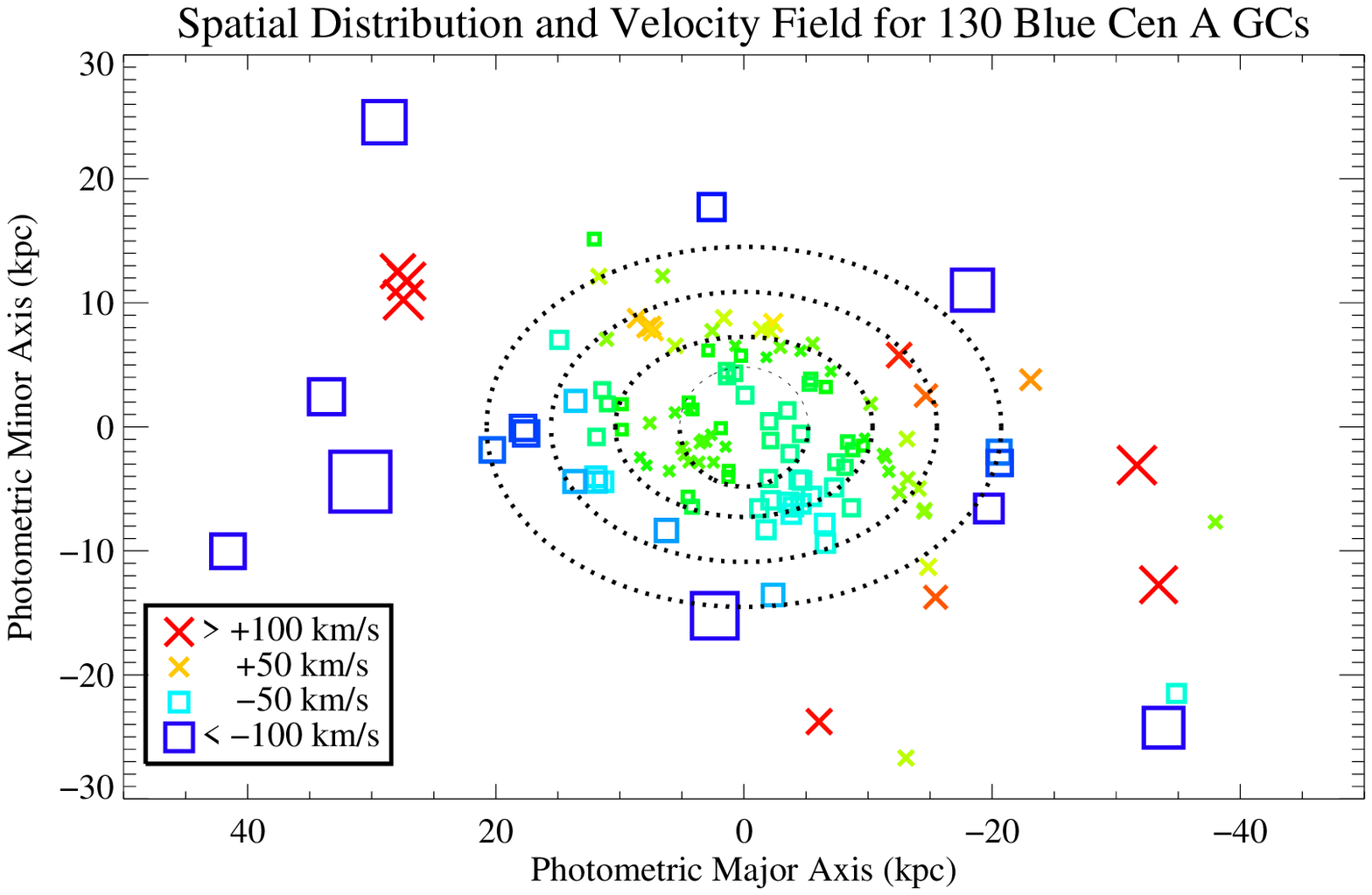}
\caption[Velocities for blue (\vi)$_0 < 1.00$ GCs]
{Velocities for blue (\vi)$_0 < 1.00$ GCs.  These figures are
same as those in Figure~\ref{figure:rvsgc_red} except
that they show the blue GCs.  There appears to be a hint of rotation,
but it is less clear than in the case of the red GCs.
\label{figure:rvsgc_blue}}
\end{figure}

As we have done before, we can also divide the sample into red and blue
GCs.  The kinematics of these sub-populations are shown in
Figures~\ref{figure:rvsgc_red} and \ref{figure:rvsgc_blue} in the same
format (actual velocities and smoothed velocities) as in
Figure~\ref{figure:rvgc_all}.  The median velocities of the two are
almost identical at 539~\kms\ for the blue GCs and 542~\kms\ for the red
GCs, both of which agree with the assumed systemic velocity of 541~\kms.
Both fields are also roughly symmetric through the origin, rejecting the
null hypothesis at the $4\sigma$\ level.

\begin{figure}
\plotone{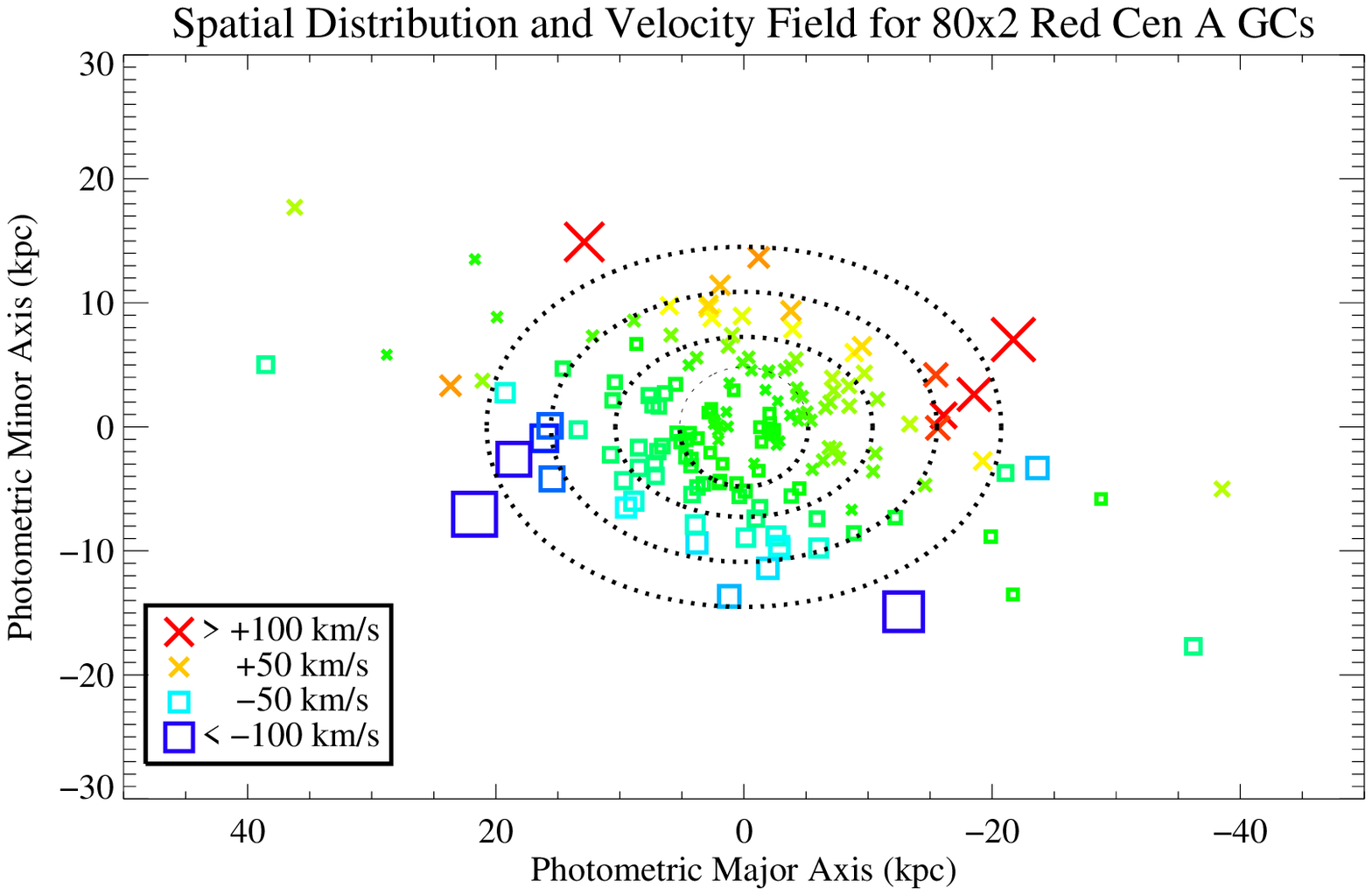}
\plotone{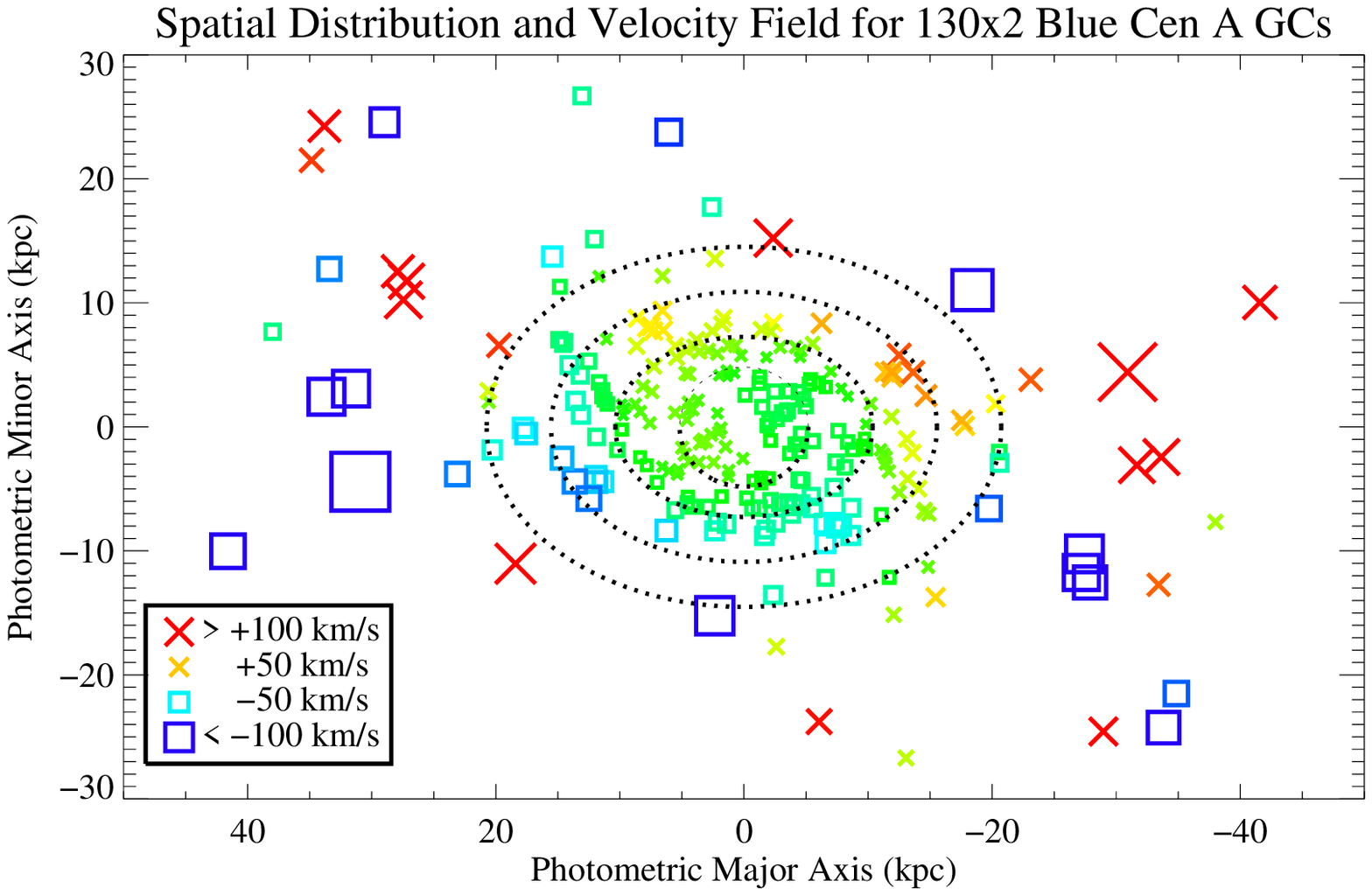}
\caption[Velocity fields for red and blue GCs]
{\small Point-symmetrized and kernel smoothed velocity fields for red
and blue GCs.  
\label{figure:rvsgc_reflect}}
\end{figure}

The smoothed velocity field of the red GCs
(Figure~\ref{figure:rvsgc_red}b) shares many characteristics with that
of the PNe, one of which is an ordered rotation about a misaligned
kinematic axis.  The velocity field of the blue GCs 
(Figure~\ref{figure:rvsgc_red}b) shows more complex structure.  
There is certainly
ordered motion, some of which is in the same sense as for the red GCs and
the PNe, but there is almost no rotation in the center and a
higher velocity dispersion in the halo.
As with the PNe, making the assumption of point symmetry in a triaxial
potential allows us to reflect each GC through the origin and
effectively double the number of velocity tracers.  The reflected and
smoothed velocity fields are shown in Figure~\ref{figure:rvsgc_reflect}.

When we do a rank-correlation analysis between the PNe and GC fields, we
find that both are positively correlated with the PNe, but the reds are
more so.
With the null hypothesis being that they are uncorrelated, we find that
the red GC velocity field rejects null at the level of $6\sigma$, while
the blue GC field rejects null at $4\sigma$.

\begin{figure}[t]
\epsscale{1.0}
\plotone{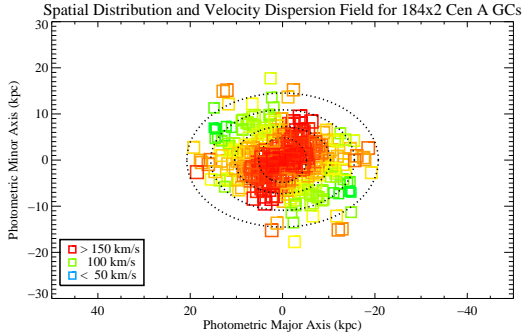}
\caption[GC velocity dispersion field]
{\small Point symmetrized velocity dispersion field for all GCs within
20~kpc.  Only in this inner region is the density of velocity tracers
high enough to determine second moment of the local velocity field.
Symbol colors represent the dynamical ``temperature'' of the stars at
the locations of each GC, with velocity dispersions ranging from 75 to
over 150~\kms.  The regions of highest dispersion are confined to a
bar-like feature that is perpendicular to the line of zero velocity
that can be seen in Figure~\ref{figure:rvsmoothgc_all}.  This is true for
both the red and blue GCs.
\label{figure:rvsgc_vdisp}}
\end{figure}

For the central regions ($R_{proj} < 20$~kpc), where the density of GCs
is high enough, we can also map the velocity dispersion field.
Figure~\ref{figure:rvsgc_vdisp} shows the point symmetrized, kernel
smoothed velocity dispersion field for all GCs within 20~kpc.  The
dispersion typically ranges from 75--150~\kms, with there being two
regions of distinct dynamical temperature.  The 
``hottest'' region is an elongated feature that lies roughly
perpendicular to the line of zero velocity.  This feature is seen in
the dispersion fields for both the red and blue GCs, and is also seen in
the PNe kinematics.

\begin{figure}
\plotone{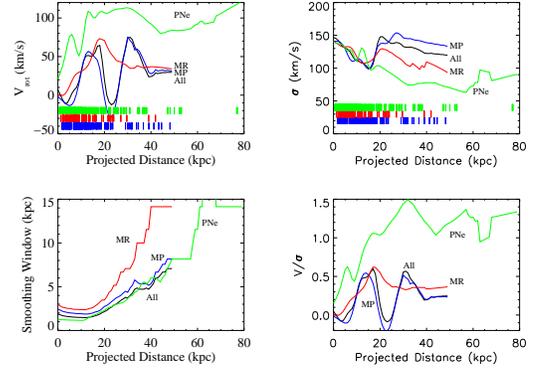}
\caption[Rotation Curves and Velocity Dispersion Profiles]
{\small Smoothed rotation curves, velocity dispersion profiles, and
smoothing scales in 
one dimension.  In all plots, the black line represents all GCs, the red
and blue lines represent the metal-rich and metal-poor GCs separately, 
and the green
line represents the PNe.  These lines are labeled as `All', `MR', `MP',
and `PNe'.  The top-left panel (a) plots the kernel smoothed
rotation curve.  The tick marks show the locations of individual objects
to give a sense of the density of the data.  From top to bottom, they
represent the PNe, MR, and MP samples.
The top-right panel (b) shows the velocity dispersion with
projected distance, and the bottom-left panel (c) shows the smoothing scale
used at a given radius.  Finally, the bottom-right panel (d) plots
$V/\sigma$ for all populations.
\label{figure:oned_kin}}
\end{figure}

Because the velocity fields exhibits some degree of symmetry, we can try
to collapse the kinematic information into one dimension.  Despite the
lack of spherical symmetry, we created a simple rotation curve and
velocity dispersion profile.  First, we changed the sign of the velocity
for all GCs on one side of the line of zero velocity.  Then, we
calculated the projected semi-major axis distance for each GC (assuming
constant ellipticity of the isophotes with radius).
Figure~\ref{figure:oned_kin} shows the resulting rotation curve and
velocity dispersion profiles against projected semi-major axis distance.
Also plotted for comparison are the smoothed rotation curve and velocity
dispersion profile for the PNe.  The last panel shows the $\sigma$ used
for the Gaussian smoothing kernel as a function of distance.

The PNe show a larger signature of rotation by a factor of two.  
This may partly be due to geometry, say if the GC system was rotation
with an inclination of $30^{\degr}$.  However, since we have no way of
determining the inclination of the GC rotation axis, we assume that we
are viewing the system ``edge-on'', as is nearly the case for the PNe.  
For the PNe, we have enough objects that we
only used PNe along the major axis to construct the one dimensional
kinematics, whereas the GCs are averaged over the entire annulus.
The PNe rotation
curve flattens at $\sim100$~\kms\ as opposed to $\sim50$~\kms\ for the
GCs.  The rotational velocity of the red GCs rises gradually to 20~kpc
before flattening.  Within 20~kpc, which is where most of the red GCs
are, the mean velocities are consistent with solid-body rotation.
The blue GCs also exhibit rotation at the level of $\sim50$~\kms, 
but again in a noisier fashion.

The velocity dispersion profile for all populations are
indistinguishable within 15~kpc.  They all have a central dispersion of
140~\kms\ that slowly declines to 100~\kms.  Outside of 15~kpc, the PNe
dispersion continues to decline to 75~\kms, while the GCs maintain a
higher dispersion.  The red GCs formally have a lower dispersion than
the blue GCs, but it is difficult to make comparisons at larger
radii because the low numbers of GCs.

\section{Mass Estimates}

Along with PNe, globular clusters are also useful tracers of a galaxy's
total mass.  GCs are luminous, discrete objects that are
relatively easy to identify.  They are also found at large radii from
the galaxy center.  Because the formation environments of GCs and field
stars may not be the same, GCs and PNe do not necessarily have similar
kinematics.  Thus, it is an important independent check to derive the
galaxy's mass from both sets of tracers.

As we discussed in our paper on the PNe (Peng \etal 2004a), 
there are numerous methods one can use to
estimate the mass.  Unfortunately, the GC system also exhibits many of
the same complexities that the PNe do.  Also, the lack of a clear
disk-like feature precludes us from limiting our kinematic analysis to
the major axis, as we did for the PNe.  As a result, we choose to
estimate the mass using the tracer mass estimator (TME; Evans \etal 2003).

The TME is optimized for the use of a ``tracer''
population, such as PN or GC, whose spatial distribution does not match
the overall mass distribution.  For this mass estimator, the mass
enclosed within the outermost tracer particle is
\[ M = \frac{C}{GN} \sum_i{v_{los_i}^2. R_i} \]
where
\[ C = \frac{4(\alpha+\gamma)}{\pi} \frac{4-\alpha-\gamma}{3-\gamma} 
\frac{1-(r_{in}/r_{out})^(3-\gamma)}{1-(r_{in}/r_{out})^(4-\alpha-\gamma)}, \]
the form of gravitational force field is $\psi(r) \propto r^{-\alpha}$,
the space density of tracer particles goes as $r^{-\gamma}$, and $r_{in}$
and $r_{out}$ define the range of radii from which the tracer objects are
drawn.  This mass estimate is only for pressure support, so the
contribution from rotation must be added.

Fitting a power law to the surface distribution of GCs, we obtained a
slope of $\gamma = 2.72$.  By assuming an isothermal halo potential
($\alpha = 0$), we can estimate the total mass of \cena\ within a given
radius.  Our sample of GCs only obeys a power law outside of 5~kpc
(because of incompleteness) so we limit our analysis to the GCs between
5 and 40~kpc.  Assuming an isotropic velocity distribution and obtaining
statistical errors via the bootstrap, we obtain a
mass within 40~kpc of $3.4\pm0.8\times10^{11} M_{\sun}$.  This is
statistically identical
to the mass of $3.5\pm0.3\times10^{11} M_{\sun}$ that we obtained from the PNe
for the same range of radii.  It is likely, however, that a significant
number of our GCs and PNe are beyond 40~kpc (Evans \etal 2003), as we
detect PNe out to 80~kpc.  If we extend the outer radius to this
distance, we obtain masses of $4.6\pm1.4\times10^{11} M_{sun}$ and
$5.3\pm0.5\times10^{11} M_{\sun}$ for the GCs and PNe, respectively.
The close agreement of these two mass
estimates suggest that the orbital anisotropy for both populations is
likely to be very similar. 

\section{Discussion: The Formation of \cena}

Ever since its association with the radio source Centaurus~A, \cena\ has
been the target of numerous studies.  Naturally, a role has often been
proposed for the radio jet in determining 
the observed optical structure --- young
stars and ionized gas along the edge of the northeast jet (e.g.\ Blanco
\etal 1975; Graham 1998; Fassett \& Graham 2000) are believed to be a
result of the interaction of the jet with dense \ion{H}{1} clouds in the
halo (Schiminovich \etal 1994).  
However, while this process surely contributes to the stellar content of
the galaxy (Rejkuba \etal 2002), it does not appear that jet-induced
star formation could have formed the bulk of the galaxy.  Rather, it is
likely that one or more merger events caused the galaxy's current
morphology, and is what supplied the gas that 
feeds the massive black hole likely to reside in the nucleus.

\subsection{An Elliptical-Spiral Merger?}

Early on, Baade and Minkowski (1954) proposed that \cena\ was the
product of a collision between an elliptical and spiral galaxy, and this
has been the traditionally favored view.  Graham (1979) used the radial
velocities of the \ion{H}{2} regions in the dust lane to show that this
feature was a rapidly rotating disk.  Malin, Quinn, \& Graham (1983) 
showed that the shells in \cena\ could be
due to phase-wrapping of stars from a captured disk galaxy with the 
size of M33.  It has also been shown 
that both the central dust lane and the \ion{H}{1} shells
can be explained by a thin, warped disk of gas and dust that
is still settling into the potential (e.g.\ van Gorkom \etal 1990, 
Sparke 1996).  We note that the \ion{H}{1}
shells in the halo have velocities that are similar to those of the PNe
in the regions that they occupy (even the one shell that crosses the
line of zero velocity). 

While the merger of an elliptical with a low-mass disk galaxy is not
ruled out by any of
the current observations, the scenario does have some problems.  
Measurements of the gas in the central disk from 
\ion{H}{2} regions (Dufour \etal 1979), and nuclear CO
absorption (Eckart \etal 1990) all indicate that it has near solar
metallicity.  This is also true for CO measurements of the gas shells
in the halo observed by Charmandaris, Combes, \& van der Hulst (2000).  
A less massive galaxy such as M33 would
have lower elemental abundances, at least initially.  Charmandaris
\etal (2000) suggest that self-enrichment from nearby star formation as
a possible mechanism to raise the metallicity in the shells, and this is
similarly a possibility in the central disk where we observe copious
ongoing star formation.  They also point out that the
amount of detectable gas in \cena\ is $\sim10^9 M_{\sun}$, a quantity
comparable to that contained in either a early-type giant spiral galaxy
or an irregular galaxy like the Large Magellanic Cloud.

Another problem with the low-mass disk merger scenario is the
stellar populations of the shells.  In the phase-wrapping scenario
proposed by Quinn (1984), a satellite galaxy falls into a rigid
potential and its stars create the over-densities in surface brightness
that we see as shells.  Therefore, if there is a substantial difference
between the stellar populations of the two galaxies, we should observe
them as color differences in the shells.  However, our color maps of
\cena\ show that, with one exception, the shells are identical in color
to the main body of the galaxy (Peng \etal 2002).  
Rejkuba \etal (2001) also find that the
color-magnitude diagram of stars in one shell is similar to that seen in
the non-shell region.  Sparke (1996) favors a dynamical age of 750~Myr
for the central disk.  If the capture happened on a similar timescale,
then the stars in the shells should still be young enough to create
detectable differences in the stellar populations.

These objections become less severe if the merger occurred with a larger
spiral galaxy, and if it occurred more than 1--2~Gyr ago.
Of course, because this scenario presupposes the existence of the
elliptical, it does not offer insight into the formation of the old
stellar component.  

\subsection{An Intermediate Age Population?}

The ages we obtain for the metal-rich GCs show that they are
significantly younger than the metal-poor GCs.  With two-thirds of the
metal-rich GCs having ages between 1 and
8~Gyr, these GCs are deemed ``intermediate age''.  There is mounting
evidence that massive metal-rich star clusters are created in starbursts,
and can survive to become GCs.  For example, Whitmore \etal (1997, 2002)
used $HST$ imaging to show that the dynamically young elliptical
NGC~3610 has a metal-rich population of GCs with an age of 4~Gyr.
Goudfrooij \etal (2001) used spectroscopy to show
that NGC~1316, a galaxy similar to \cena, has 3~Gyr old GCs that are
believed to be result of a recent merger.  Forbes \etal (2001) also
found evidence for possible intermediate age GCs in NGC~1399.
Puzia (2002) studied
the GC systems of a number of ellipticals and found that
while metal-poor GCs are almost uniformly old, metal-rich GCs show a
large spread in ages.  However, not all metal-rich GCs are necessarily
young.  Studies of other GC systems, such as M87 (Jord{\'a}n \etal
2002), M49 (Puzia \etal 1999),
and the Galaxy, show the metal-poor and metal-rich GCs to be nearly
coeval.

In \cena\ itself, there are many lines of evidence that point to the
presence of an intermediate-age population.  Alonzo \& Minniti (1997)
used optical-infrared photometry to determine that there are a number of
``IR-enhanced'' GCs in the central 3~kpc.  These clusters have redder
colors due to the presence of carbon stars along the asymptotic giant
branch (AGB), a feature of stellar populations with ages of 1--4~Gyr
(Aaronson \& Mould 1982).

There is also evidence that the field star population has an
intermediate age contribution.  Using optical-IR color-magnitude
diagrams, Rejkuba \etal (2001) found that there are a significant
number of bright AGB stars in the field population --- up to 10\% of the
field stars may be intermediate-age.  Soria \etal (1996) and
Marleau \etal (2000) came to similar conclusions using $HST$ WFPC2 and
NICMOS observations.  This, however, is at odds
with the $HST$ WFPC2 work of Harris, Harris, \& Poole (1999; HHP99) and 
Harris \& Harris (2000; HH00) who argue that the
contribution of intermediate-age stars to the halo of \cena\ is no more
than 1\%.  It is possible that these discrepencies can be accounted for
by a radial gradient in age as the HHP99 and HH00 fields are at 21 and
31~kpc, while the fields for the other studies are at radii
between 8 and 14~kpc.  There may be other spatial dependencies as well,
depending on whether fields are observed along the major axis and the
extended stellar disk --- one Rejkuba \etal (2001) field was on the
major axis, but both HHP99/HH00 fields were chosen to avoid it.  

The integrated stellar populations of ellipticals also show a spread in
ages. Trager \etal (2000) used Lick spectral indices to show that while
cluster ellipticals were generally old, field ellipticals had a younger
mean age with a spread of ages, modulo the contribution of metal-poor
BHB stars (Maraston \& Thomas 2000).  
This is consistent with a picture where
interactions and mergers drive evolution, and where these processes,
especially gaseous interactions, happen later in less dense environments.
As a galaxy in a loose group,
\cena\ would be considered a field elliptical.

Our observations of the metal-rich GCs and the observations others have
made of the field star population support the conclusion that a
significant fraction of the stars in \cena\ have intermediate ages.  
GCs systems are generally less centrally concentrated than their
corresponding field stars, and it is possible that most intermediate-age
field stars are either in the central regions, or perhaps localized
along the major axis.

\subsection{Multi-phase {\it In Situ} Formation}

In some ways, the simplest scenario of galaxy formation is the monolithic
collapse of a protogalactic gas cloud (Eggen, Lynden-Bell, \& Sandage
1969).  Forbes, Brodie, \& Grillmair (1997) proposed a ``multi-phase''
collapse scenario in order to account for the bimodal metallicity
distributions of GC systems.  In this scheme, the initial formation
epoch is truncated, perhaps by supernova winds, and the subsequent
formation episodes occur when the enriched gas can again form stars.

When trying to recreate the metallicity distribution function (MDF) of
field stars in \cena, HH00 and HH02 found it necessary to invoke two-phases
of chemical evolution.  At early times, they used an
``accreting-box'' model which allowed for gas inflow, and at late times,
they treated the galaxy as a ``closed-box'', shutting off any gas
inflow.  Like in the Solar neighborhood, gas inflow is necessary to
avoid an overabundance of metal-poor stars as compared to observations
(the classic ``G-dwarf problem'').  In spirit, this is similar to the
scenario proposed by Forbes \etal (1997), 
although it is not exclusive to their scenario.  

In a multi-phase collapse scenario, one would expect that the
metal-rich GCs and the PNe (field stars) would have similar kinematics
since they were formed out of the same gas.  The metal-poor GCs, having
very few corresponding field stars, would not necessarily have
correlated kinematics.  While our data agrees with this in principle,
we believe that a larger issue is not the agreement or disagreement between
metal-poor GCs, metal-rich GCs, and PNe, but the fact that all three
populations exhibit a significant amount of angular momentum in their
halos.  This is more likely to be a result of angular momentum
transport than of a rapid dissipational collapse.
In addition, in the case of \cena, the delay between the two epochs of
star formation must be many billion years.  Because the timescale for 
the return of supernova heated gas to the cool ISM is still uncertain, this
length of a delay may yet be plausible.
Nevertheless, while a multi-phase {\it in situ} collapse may be a sufficient
description of the galaxy's chemical 
enrichment history, it will be important to determine a set of
corresponding kinematic predictions.

\subsection{\cena\ as a Result of a Disk-Disk Merger}

The mergers of disk galaxies have been suggested as a possible formation
mechanism for ellipticals for over thirty years (Toomre \& Toomre
1972).  A classic objection to this model was that ellipticals have many more
GCs per unit light ($S_N$) than spirals, and that this would also be
true for the sum of two spirals.  
Ashman \& Zepf (1992) proposed that disk-disk mergers could produce new
globular clusters that would make up for the perceived deficit of GCs in
spirals per unit light as compared to ellipticals.  Whether or not these
newly formed GCs can actually raise the specific frequency of the
resulting galaxy is still debated, but
both observations and simulations show that these mergers can
result in intense starbursts that eventually settle into hot stellar 
systems (Schweizer 1996, Barnes \& Hernquist 1996). 
Hernquist \& Spergel (1992) also show that major mergers can 
reproduce the shells
seen in some ellipticals, and that leftover gas can
re-accrete into a warped, rotating central gas disk (Mihos, private
communication).  

N-body simulations of disk-disk mergers also show that these can produce
a wide variety in dynamical characteristics in the remnant.  
Equal-mass (1:1) mergers are likely to produce boxy, non-rotating remnants,
while unequal-mass (3:1) mergers produce disky, rotating remnants (Bendo \&
Barnes 2000; Naab \& Burkert 2001).  The latter result is similar to
what we observe in the spatial distribution and kinematics of the PNe.

Recently, Bekki, Harris, and Harris (2003; BHH03) simulated a disk-disk merger
(without gas) in an effort to reproduce the \cena\ field star MDF in the
outer halo.  They found that most of the stars in the halo of their
merger remnant originate in the {\it outer disks} of the progenitors, 
helping to explain why the observed MDF has a preponderance of
relatively metal-rich stars.  These disk stars gain orbital
angular momentum from their parent galaxies and become more spatially
extended, mixing with the traditional metal-poor stellar halo.  
By contrast, contributions from the
bulges or any gas dissipation are confined to the central regions of the
remnant.  

On many accounts, the scenario by which at least the metal-rich
component of \cena\ formed 
via a disk-disk merger is attractive.  In this merger of
two spirals of unequal mass, the larger galaxy's angular
momentum vector is parallel to the current photometric minor axis, with
the extended stellar disk along the major axis being a remnant of this
system.  The smaller galaxy may have had an angular momentum
vector perpendicular to the larger galaxy's, with its stars following
the x-tube orbits of the combined potential.  An analysis of the H95 PNe
data by Mathieu, Dejonghe, \& Hui (1996) estimated that 75\% of the PNe
were on z-tube orbits (minor axis rotation) with the other 25\% on
x-tubes, a ratio of 3:1.  
This event, which occurred 3--8 Gyr ago, formed the bulk of the
metal-rich GCs, and the gas that did not form stars, which would
presumably have nearly solar abundance, accumulated in 
the central gas disk and \ion{H}{1} shells.  Forming out of gas that
originally shared its kinematics with the progenitors' disk stars, the
metal-rich GCs then rotate in much the same way as the PNe.  And while
the metal-poor GCs may receive some angular momentum from the merger,
they do not rotate as much as the PNe or metal-rich GCs because they
never had as much initial angular momentum.

Problems with this scenario arise when one examines the number and spatial
distributions of the GCs.  BHH03 point out that their simulation
provides no mechanism for metal-rich GCs to exist in the halo, only in
the central regions.  Yet, we do see a number of metal-rich GCs beyond
$2r_e$.  This is likely more a problem with the simulations, as the
treatment of gas cooling and reheating is important for predicting the
final spatial distribution of clusters.
Another problem is that with over 60\% of the known GCs
classified as metal-poor, there are too many metal-poor GCs in \cena\ to have
originated solely from the halo populations of two spirals such as the
Milky Way and M31.  Forbes \etal (2000) calculated
that a dissipationless merger of all Local Group galaxies would
eventually produce 
a faded remnant with an absolute $V$ magnitude of $M_V = -20.9$ and have
$\sim700$~GCs.  When compared to \cena ($M_V = -21.9$, Dufour \etal
1979, H95), the luminosity does not pose a problem if the merger remnant
only fades for 5~Gyr, although the total dynamical mass of the remnant
would be larger (Evans \etal 2003).  Also, the relatively low fraction
of old, metal-rich GCs points to the possibility that the progenitor
disks had relatively small bulges (and their associated bulge GC
systems).  Despite these issues, though, a disk-disk merger is a promising
framework within which we can explain many of \cena's observed properties.

\subsection{\cena\ as a Result of Hierarchical Merging}

There is already ample evidence that \cena\ experienced one or many
mergers in its recent past.  For example, the young tidal stream is
a resident example of the accretion of low mass objects.  Hierarchical
models of galaxy formation (e.g.\ Kauffmann, White, \& Guiderdoni 1993) 
predict that
merging should be commonplace in the universe, and should be one of the
main drivers of galaxy evolution.  C\^{o}t\'{e}, Marzke, \& West
(1998) showed that hierarchical mergers involving little gas dissipation
could still produce the bimodal metallicity distribution seen in GC
systems.  Using a semi-analytic galaxy formation prescription, 
Beasley \etal (2002) simulated the GC systems of ellipticals, including
both gas-poor and gas-rich merging, and also found them to have bimodal
metallicity distributions.
The metal-rich GCs in Beasley \etal (2002) model are the result of
gaseous merging, most of which happens between
redshifts of 1 and 4 (8--12~Gyr ago in the standard cosmology), 
and have mean ages that are younger than the mean ages
of the metal-poor GCs.  
These models are attractive in that they
combine aspects of all the previously discussed scenarios.

Beasley \etal (2003) also explicitly applied their semi-analytic models
to the task of reproducing the metallicity distribution functions of the
field stars and GCs in \cena.  While most of the star formation in their
model galaxies happens early, they do predict a relation between age and
metallicity in the GCs, with many of the metal-rich GCs forming in
merger-induced ``bursts''.  While a large number of intermediate-age GCs
are not inconsistent with their models --- in fact, some of their model
galaxies experience significant bursts at late times --- the existence
of a single intermediate-age metal-rich population would be a rare
event and inconsistent with their age-metallicity relation.  The current
data do not allow us to distinguish between these two possibilities, but
higher quality data can be used to derive the age distribution of GCs
and constrain this aspect of these models.

By viewing the formation of \cena\ in the context of multiple mergers,
we can reconcile some of the issues we have raised.  The relatively late
gaseous merger ($z\sim0.5$) as compared to the Beasley \etal (2002)
models can be attributed to the low density environment of \cena.  In
the group environment, all merging is expected to happen later.
The deficit of metal-poor GCs in a disk-disk merger can be accounted for
by the accretion of dwarf elliptical (dE) galaxies.
HH00 claim that because the halo MDF is predominantly metal-rich, 
it could not have been formed by accretion of dEs.
However, it is only the metal-poor tail of the MDF that needs to
originate from the dEs, not the entire MDF of the halo.  HHP99 calculate
the ``metal-poor $S_N$'' is 4.3, which is very much in line within the
values of 3--7 for dEs in Virgo (Miller \etal 1998).  Accretion of dIs,
such as we see happening with the young tidal stream, can also
contribute metal-poor GCs, but they have specific frequencies that are
too low to have made up the bulk of the metal-poor halo.

\begin{figure}[t]
\epsscale{1.0}\plotone{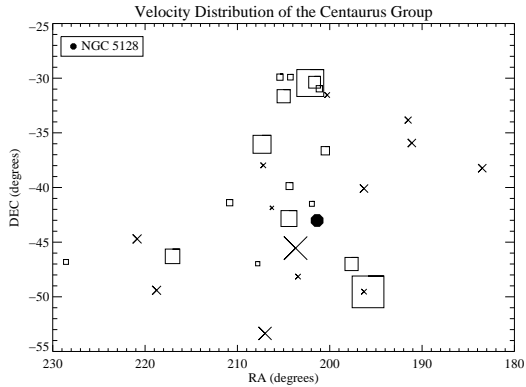}
\caption[Spatial Distribution and Velocity Field of the Centaurus Group]
{\small Spatial distribution and velocity field of the Centaurus Group.
\cena\ is 
represented by a filled circle.  All other galaxies are marked by their
position and velocities with respect to 541~\kms.  The largest symbols
are $\pm200$~\kms\ from the systemic radial velocity of \cena.  There
does not appear to be any obvious rotation in the group.  Data taken
from Banks \etal (1999).
\label{figure:cenagroup}}
\end{figure}

If the metal-poor GCs are accreted, then how can they also show some
co-rotation with the stars?  Bekki \etal (2002) claim that any GCs that
exist before a merger will gain angular momentum through dynamical
friction.  As for the ones that accrete later, they
may also have preexisting angular momentum if the group environment
itself rotates. 
C\^{o}t\'{e} \etal (2001) found that the metal-poor GCs in M87 rotate
as a solid body about the same axis as the cluster angular momentum.
Thus it is also important to examine the kinematic environment of \cena.
Figure~\ref{figure:cenagroup} shows the spatial and velocity
distribution of galaxies in the Centaurus group with respect to \cena\ 
(data taken from Banks \etal 1999).  There does not appear to be any
obvious rotation of the group as a whole, but this does not necessarily
present a problem, as observations show that the metal-poor GCs in fact
have a higher velocity dispersion than the rest of the galaxy.  This is
perhaps due to the continuing late accretion of metal-poor systems.

\section{Conclusions}

While one must always be cautious about trying to infer too much from a
sample of one galaxy, detailed studies of \cena\ can teach us much
about the formation of giant elliptical galaxies.  As the nearest gE
galaxy, radio galaxy, and recent large merger, \cena\ is an accessible
laboratory where we can study 
the processes that likely played important roles forming most elliptical
galaxies at higher redshift.

The stellar populations and kinematics of \cena\ show its history to be
consistent with a hierarchical scenario in which most
metal-poor GCs formed early and merged with the galaxy or its
progenitors over time.  The major event that shaped the galaxy as we see
it today was likely a disk-disk merger, perhaps with unequal masses,
that produced a disky, rapidly rotating remnant with a 
metal-rich GC population.  This occurred at
intermediate times ($\sim5$~Gyr ago, or $z\sim0.5$), and may also be
responsible for the gas currently in the galaxy, although a subsequent
merger of another gas-rich system with the newly-formed elliptical is
not ruled out.  The gas that resulted from this merger currently feeds
the active nucleus, which in turn triggers star formation in the halo.

Future work that increases the number of known GCs, and obtains better ages
and metallicities for all GCs will be important for corroborating or
extending this formation picture.  As well, scenarios and
simulations that provide chemodynamical predictions will be essential
for making full use of these kinds of data sets.
Ellipticals are likely to have diverse formation histories that are a
function of their mass and environment.  It will be desirable to extend
detailed studies of GC and PN systems to a larger sample of galaxies in the
nearby universe.

\acknowledgments

E.\ W.\ P.\ acknowledges support from NSF grant AST 00-98566. 
H.\ C.\ F.\ acknowledges support from NASA contract NAS 5-32865
and NASA grant NAG 5-7697.  We thank David Malin for
making available to us his deep photographic prints of NGC 5128. 
We also thank Brad Whitmore for making available to us the KMM code.
This research has made use of the NASA/IPAC Extragalactic
Database (NED), which is operated by the Jet Propulsion Laboratory,
California Institute of Technology, under contract with NASA.

\end{document}